\documentclass{article}
\usepackage{amssymb}

\begin{document}

\newtheorem{defi}{Definition}

\title{\bf{A Cosmological Sector in Loop Quantum Gravity}}
\author{Tim A. Koslowski\\
\texttt{tim@physik.uni-wuerzburg.de}\\
Institut f\"ur Theoretische Physik und Astrophysik\\
Universit\"at W\"urzburg\\
Am Hubland\\
97074 W\"urzburg\\
Germany, European Union}
\date{\today}
\maketitle
\thispagestyle{empty}

\begin{abstract}
  We use the method of embedding a subsystem (i.e. its observable algebra) into a larger quantum system to extract a cosmological sector from full Loop Quantum Gravity. The application of this method provides a setting for a systematic study of the interplay between diffeomorphism invariance and symmetry reduction. The non-triviality of this relation is shown by extracting a cosmological system that has configurations variables that are very similar to the ones of a super-selection sector of standard Loop Quantum Cosmology. The full operator algebra however turns out to be different from standard Loop Quantum Cosmology. The homogeneous isotropic sector of pure gravity turns out to be quantum mechanics on a circle. The dynamics of our system seems pathological at first sight, and we give both mathematical and physical reasons for this behavior and we explain a strategy to cure these pathologies.
\end{abstract}

\section{Introduction}

Cosmological models are one of the most important areas of application for General Relativity. This is on the one hand due to the importance of gravity in the description of the dynamics of the universe as a whole, but also due to the huge simplifications that occur mathematically, when one considers homogeneous models that describe large scale cosmology, which make exact calculations possible, so one can make strong and unambiguous predictions. The classical cosmological models are indeed sectors of General Relativity in the sense that they can be embedded into full General Relativity or conversely that one can reduce the phase space of General Relativity to the respective mini-super-phase space.

The importance of quantum gravity in the early universe has been stressed for a long time. It is thought to be responsible for providing initial conditions and a consistent dynamics of the universe very close to the classical big bang, particularly it is expected to resolve singularities. A successful theory of quantum gravity is provided by Loop Quantum Gravity (LQG) \cite{rovelli,thiemann} which has sparked investigations into Loop Quantum Cosmology (LQC) premiered by Bojowald \cite{bojo}. LQC is a symmetry reduction of LQG and shares many features of the full theory of Loop Quantum Gravity, however the full relation between Bojowalds LQC and full LQG is not completely understood. LQC can be constructed using symmetric states on a lattice and then averaging the fundamental operators over the isometry group \cite{bojo-inhom,bojo-review}. There are however concerns \cite{engle-qft} that this implementation of symmetry does not precisely capture the symmetric sector of LQG. It is the aim of this paper is to provide a general construction principle for symmetric sectors in LQG and to construct cosmological sectors thereof.

Recently we introduced a general concept of reducing a quantum theory of a system to a subsystem that does not require a classical reduction \cite{reduction}. This procedure is based on the analogue of the pull-back under a Poisson-embedding of observables from a full classical theory to a reduced classical theory. We will work with an adaption of the Weyl-algebra setting for LQG \cite{fleischhack-weyl} in this paper, which allows us to work in the the $C^*$-algebraic framework. This provides in principle a precise prescription for the reduction of composite operators, particularly for those that are elements of the $C^*$-algebra. A certain class of $C^*$-algebras and certain classical embeddings, that we called ''full'' in \cite{reduction} can be turned into quantum embeddings. These qunatum embeddings can be used to construct the $C^*$-algebra of the reduced system directly from the full theory for which they induce a Hilbert space representation out of a given Hilbert space representation of the full quantum algebra.

This approach to the construction of a reduced quantum system is applied to Loop Quantum Gravity in this paper. We start with the classical Bianchi I sector in Ashtekar variables and construct a corresponding quantum embedding. This needs a gauge fixing of the diffeomorphisms, which we achieve for a set of extended diffeomorphisms whose physical significance where explained in \cite{diffeo}. This set of diffeomorphisms acts most naturally as a groupoid as we will show in the course of this paper. It turns out that the construction of the quantum embedding is not free of ambiguities, although the classical embedding is unique. We construct a ''simplest'' embedding and discuss briefly other possibilities. This quantum embedding is then used to extract a cosmological sector from the full theory of Loop Quantum Gravity. As a first step, we extract the locally rotationally symmetric and the isotropic sector of pure gravity on the base manifold $\sigma=\mathbb R^3$. The isotropic sector turns out to be equivalent to usual quantum mechanics of a particle on a circle and the locally rotationally symmetric sector turns out to be quantum mechanics on a 2-torus. Both systems are represented on a separable Hilbert space. Thereafter we give an example for the treatment of Loop Quantum Gravity with matter within our approach.

Using the fact that Gravity is a theory with a constrained Hamiltonian we give a naive construction of the dynamics for this system as it is induced by the full theory. This dynamics is however not compatible with the classical system, because it has too many solutions to the Hamilton constraint, which is due to the over-representation of states that correspond to degenerate geometries in the full theory. We explain how the construction of the quantum embedding can be refined to get rid of this over-representation of degenerate geometries, which is based on the physical requirement of geometrical homogeneity of the states used in our construction.

It turns out, as we show in the discussion of the differences of our approach to Loop Quantum Cosmology, that the flux operators of the reduced sector have a countable spectrum, which differs from Loop Quantum Cosmology. There is also no $\mu_o$ parameter in our system. These differences can be attributed to the different fixing of the diffeomorphisms compared to standard LQC. Once the dynamics of our system is constructed, one can however hope to fix the $\mu_o$ ambiguities in the dynamics of Loop Quantum Cosmology, such that the dynamics of our cosmological sector and the dynamics of Loop Quantum Cosmology coincide.

This paper is organized as follows: We review the necessary preliminaries in section 2. This includes a brief introduction of Bianchi Cosmology, mainly to fix the notation of this paper, a review of our methods of reduction of quantum theories using quantum embeddings, as they where introduced in \cite{reduction}, an a brief review of a combinatorial approach to Loop Quantum Gravity that turns out to be useful for this paper. The details about this approach will be published in a future paper. We construct a quantum embedding map for cosmology in section 3, which we use in section 4 to extract a cosmological sector from full Loop Quantum Gravity. Since this sector differs from standard Loop Quantum Cosmology, we give a discussion of these differences in section 5.

\section{Preliminaries}

This section contains preliminaries on Bianchi cosmology, quantum embeddings and a combinatorial approach to LQG to make this paper self-contained and to fix the notation used throughout this paper.

\subsection{Bianchi Cosmology}
\label{red-conn}

Bianchi type cosmologies are standard, but since we need the explicit formulae for Ashtekar variables for these models, we include a small discussion in this paper. Assuming a 3-dimensional spatial base manifold $\sigma$ and a group $\mathbb G$ of isometries, i.e. a free and transitive action of a group of diffeomorphisms $\phi$ on $\sigma$ for which we assume $\phi^* q=q$ for the spatial metric $q$ on $\sigma$. Denoting the Killing vector-fields that generate $\mathbb G$ by $X$, we obtain a Lie-algebra $[X_i,X_j]=C^k_{ij}X_k$. Using triads $e^a$, that commute with the Killing vector fields, we use the unit normal $n$ to $\sigma$ and bundle these together to a quadruple $e_\mu$, whose integrals furnish local coordinates. Since the commutation relations $[e_\alpha,e_\beta]=\gamma^\mu_{\alpha\beta} e_\beta$ are functions independent of the local spatial coordinates, we can expand the structure constants as $\gamma^i_{jk}=\epsilon_{jkl} n^{il}+a_j\delta^i_k+a_k\delta^i_j$, where $a$ and $n$ are constants in the local frame satisfying $n^{ij}a_j=0$. We are interested in Bianchi-type A cosmologies i.e. $a_i=0$ (no shift) and there in particular in type I cosmology, i.e. $n_i=0$ in the local frame. In this case, we obtain a base manifold that is $\sigma=\mathbb R^3$ with a spatial metric $q$ that is constant in local coordinates.\footnote{This paper is concerned with Bianchi type I cosmology only.}

We will increase the symmetry of our model further by assuming local rotational symmetry (i.e. rotational symmetry around one fixed axis) or isotropy (i.e. rotational symmetry around all axes, which can be achieved by the introduction of additional vector fields $X_i$ which take the form $Y_i=\epsilon_{ijk}x_j\partial_k$ in the local frame in which $X_i=\partial_i$.

A $\mathbb G$-symmetric connection on a fibre bundle over $\sigma$ can be decomposed into a reduced connection on a reduced bundle over a reduced space $\sigma/\mathbb G$ and a $\mathbb G$-multiplet of scalars on the reduced space. This stems from the fact that connections are by definition invariant under vertical bundle morphisms (i.e. gauge transformations) and the following observations \cite{brodbeck}: Consider a Lie-group $\mathbb G$ acting as a group of bundle morphisms on a principle fibre bundle $P(\sigma,\mathbb H,\pi)$, such that all $\mathbb G$-orbits are isomorphic, $\sigma/\mathbb G$ is reductive and $\mathbb I$ is the isotropy group of a point, such that $\sigma=(\sigma/\mathbb G)\times(\mathbb G\times \mathbb I)$, i.e. $\sigma$ is the orbit bundle over $\sigma/\mathbb G$. Since each point $p \in P$ defines a morphism $\rho_p:\mathbb I \rightarrow \mathbb H: i \mapsto \alpha_i(p)$, where the action $\alpha$ commutes with the right action in the fibre, yielding: $\rho_{ph}=Ad_{h^{-1}}\rho_p$. Fixing one particular $\rho$ allows the construction of the subbundle $P_{sym}(\sigma/\mathbb G,C_{\mathbb H}(\rho(\mathbb I)),\pi|_{P_{sym}})$ with the reduced structure group given by the centralizer of the $\rho$-image of $\mathbb I$, which is isomorphic to any subbundle built from $\rho^\prime=Ad_{h^{-1}}\rho$ in the conjugacy class of $\rho$, such that the $\mathbb G$-symmetric fibre bundle is completely classified by $(P_{sym},[\rho]_{conj})$.

A $\mathbb G$-symmetric connection on a principle fibre bundle $(P_{sym},[\rho]_{conj})$ defines (1) a connection $\omega_{sym}$ by restriction and (2) a linear map $L_p:\mathfrak G \rightarrow \mathfrak H : V \mapsto \omega_p(V)$ at each point $p$. The $L$-image of the orthogonal complement $\mathfrak I_\perp$ of $\mathfrak I \subset \mathfrak G$ is horizontal but not tangential to $\sigma/\mathbb G$. Invariance under vertical bundle morphisms implies $L_p(Ad_iV)=Ad_{\rho(i)}(L_p(V))$, defining the transformation law for the components $L_p|_{\mathfrak I_\perp}$, which defines the transformation law for the scalar multiplet $L$. Using the Maurer-Cartan form $\theta_{MC}$ on $\mathbb G$ and the embedding $i:\mathbb G/\mathbb I \rightarrow \mathbb G$ we can write the $\mathbb G$-symmetric connections as $\omega=\omega_{sym}+L\circ i^* \theta_{MC}$, where $\omega_{sym}$ is a connection on $P_{sym}$.

This framework was applied to various symmetric models \cite{bojo-kast} and in particular to Bianchi type I cosmologies \cite{bojo}. The topology of the base manifold is $\sigma=\mathbb R^3$ upon which the symmetry group $\mathbb G=\mathbb R^3$, generated by three generators $g_i=\partial_i$ in the local frame, acts as translations. The left-invariant one-forms on $\mathbb G$ take the form $\omega^i=dx^i$, which allows us to express $\theta_{MC}=g_i dx^i$. The isotropy group $\mathbb I$ is trivial s.t. there is only the identity embedding $i:\mathbb G/\mathbb I=\mathbb G \rightarrow \mathbb G$. The reduced space $\sigma/\mathbb G=\{x_o\}$ consists of one point $x_o$ only as does the reduced principal fibre bundle. The linear map $L$ becomes a matrix over this point, such that the symmetric connection is:
\begin{equation}
  A = L\circ \theta_{MC}=L^I_i \tau_I \omega^i=A^I_a \tau_I dx^a,
\end{equation}
where $\tau_I$ denotes the $i$th generator of $SU(2)$. The transformation constraint under vertical bundle morphisms becomes trivial and is thus satisfied by all matrices $L$. Using the left-invariant vector fields $\omega^i(X_j)=\delta^i_j$ allows us to write the Ashtekar variables as:
\begin{equation}
  A^I_a=l^I_i \omega^i_a \textrm{ and } E^a_I =\sqrt{|g|} m^I_I X^a_i,
\end{equation}
where $g$ is the invariant metric $g=\delta_{ij}\omega^i\vee\omega^j$.

Imposing additional local rotational symmetry around the $3$-axis amounts to enlarging the isotropy group $\mathbb I$. The embeddings of $\rho_n:\mathbb I=U(1)\rightarrow \mathbb H=SU(2): \exp(a Y) \mapsto \exp(n a \tau_3)$ are labeled by integers $n$ which satisfy the transformation constraint only for $n=1$ with matrices of the form:
\begin{equation}
  L=\biggl(
    \begin{array}{ccc}
      a&b&0\\
      -b&a&0\\
      0&0&c
    \end{array}
    \biggr),
\end{equation}
for three real numbers $a,b,c$.

Enlarging the symmetry group further to isotropic connections enlarges the isotropy group to $\mathbb I=SU(2)$ which can be embedded by the identity map into $\mathbb H=SU(2)$. The solutions to the transformation equations yield the matrices of the form
\begin{equation}
  L\,=\,c \,\,\mathbb I_{3\times 3}.
\end{equation}
We notice that the parameters $a,b,c$ are the degrees of freedom of the LRS-model, however we are always able to go into a diagonal gauge in which the parameter $b$ vanishes.

For our discussions of cosmological models it will be important that one can give a closed formula for the holonomy of a homogeneous connection along a straight edge\footnote{The attribute straight means here a straight line in the local frame in which the connection components are constant matrices $A(x)=A^i_a dx^a \tau_i$.}.
Given the matrix $A$ containing the elements of $A^i_a$ and given an edge that can be expressed in this chart as:
\begin{equation}
  e=\{e^a(0)+l \hat{e}^a t: 0 \le t \le 1\}
\end{equation}
then we can calculate the holonomy along this edge
\begin{equation}
  h_e(A)=\mathcal P\biggl\{\exp\biggl(l \int_0^1 dt A^i_a \hat{e}^a \tau_i\biggr)\biggr\}
\end{equation}
explicitly due to the simplifications arising from (1) the fact that the $A^i_a$ are constant, making the path ordering trivial and (2) the explicit form of the $\tau^i$, which allows us to calculate the Lie group exponential function as a matrix exponential function.

\subsection{Reduction of Quantum Theories}

We presented a construction in \cite{reduction} that allows the explicit reduction of certain quantum systems in a way that is a noncommutative analogue to the pull-back under a Poisson-embedding in a classical system. This construction can be applied to quantum systems that are given by the following data: $(\mathfrak A(\mathbb X,\mathbb G),\pi,\mathcal H)$. $\mathbb X$ is a locally compact Hausdorff space that serves as the quantum configuration space, s.t the configuration variables arise as $C(\mathbb X)$. $\mathbb G$ is a group of Weyl-operators (stemming from exponentiated Poisson actions of momenta) whose action $\alpha$ is ''unitary'' on $C(\mathbb X)$ and acts freely and properly as pull-backs under homeomorphisms of $\mathbb X\rightarrow \mathbb X$. $\mathfrak A(\mathbb X,\mathbb G)$ denotes the crossed product $C^*$-algebra of $C(\mathbb X)$ and $\mathbb G$ by $\alpha$, whose product is given by the convolution product $f_1*f_2(x,g):=\int d\mu_H(h) f_1(x,h)f_2(\alpha_{h^{-1}},h^{-1}g)$. $\mathcal H$ is a Hilbert-space carrying a regular representation $\pi$ of $\mathfrak A$.

The particular classical Poisson embeddings $\eta:(\mathbb X_o,\Gamma_o)\rightarrow(\mathbb X,\Gamma)$\footnote{The pair $(\mathbb X_i,\Gamma_i)$ denotes a classical phase space $\Gamma_i$, that has a distinct subset $\mathbb X_i$ that serves as the classical configuration space.} that we considered where ''full'' in the sense that all momenta that are tangential to the subspace $\eta(\mathbb X_o)\subset\mathbb X$ are supposed to be in $\Gamma_o$. Given a classical system, whose quantization is given by $(\mathfrak A(\mathbb X,\mathbb G),\pi,\mathcal H)$ and a full  embedding $\eta$ of a classical subsystem into the full classical system raises the question: ''What are the compatible quantizations of this subsystem, in the sense of preserving the algebraic structure and the expectation values of the full quantum system?'' This exactly the question for ''providing a quantum subsystem'' or mathematically for ''constructing a noncommutative embedding''.

The basic observation behind this construction is given in terms of special pre-Hilbert-$C^*$-modules, which themselves have a commutative algebraic structure. For the $C^*$-algebras $\mathfrak A(\mathbb X,\mathbb G)$, there are ''canonical'' pre-Hilbert-$C^*$-modules given by $C_c(\mathbb X)$ together with the bilinear form:
\begin{equation}
  \langle f_1, f_2 \rangle_A : h \mapsto f_1(x) \int d\mu_H(g) \overline{f_2(\alpha_{g^-1}x)} h(\alpha_{g^{-1}}x),
\end{equation}
which is dense in $\mathfrak A$. The observation that full embeddings are determined by embeddings of the configuration space suggests to consider the restriction $\eta|_{\mathbb X}$ to the embedding of the configuration space. This restriction defines a map $P$ on $C(\mathbb X)$ by:
\begin{equation}
  P: C(\mathbb X) \rightarrow C(\mathbb X_o): f \mapsto \eta|_{\mathbb X}^* f,
\end{equation}
which is a linear map form the linear space of configuration observables on $\mathbb X$ into equivalence classes of observables that coincide at the embedding $\eta(\mathbb X_o)$. By the axiom of choice, there exists linear inverses $i$ that map each equivalence class into a representative in this equivalence class such that the entire map $i$ is again linear. Such a pair $(P,i)$ of linear maps contains all information about a full embedding $\eta$, and we called it a quantum embedding, because using them we can provide the following construction: We can consider a reduced linear space $P(C_c(\mathbb X))$, that serves as a module over the induced ''rank-one'' operators
\begin{equation}
  \langle P(f_1),P(f_2) \rangle_{red} : P(h) \mapsto P(\langle i(P(f_1)),i(P(f_2))\rangle_A i(h)),
\end{equation}
which turns out to well defined. Moreover, one can use standard techniques to show that these induced rank one operators span a Weyl-algebra that arises as the quantization of a fully embedded reduced quantization. In analogy to Rieffel induction, one can also induce a Hilbert space representation for the embedded Weyl-algebra out of a given Hilbert space representation of the full system. Any representation can be written as a direct sum of cyclic representations $(\mathcal H_\omega,\pi_\omega)$ out of states $\omega$. The states $\omega$ on the full theory induce states $\omega_{red}$ on the reduced theory by extending
\begin{equation}
  \omega_{red}(\langle P(f_1),P(f_2)\rangle_{red}):=\omega(\langle i(P(f_1)),i(P(f_2))\rangle_A)
\end{equation}
by density to the reduced Weyl-algebra. The induced representation is then given by the direct sum of the GNS-representations out of the induced states $\omega_{red}$.

A very useful tool for constructing correspondences of operators of the full theory with operators in the reduced theory is given by an approximate identity $id_{C,U(e),\epsilon}(x,g)$, labeled by increasing compact subsets $C \subset \mathbb X$, shrinking open neighborhoods $U(e)\subset \mathbb G$ of the group identity element $e$ and a shrinking real number $\epsilon>0$, which has the following properties:
\begin{equation}
   \begin{array}{rcl}
     id_{C,U(e),\epsilon} & = & 0 \, \textrm{ for } g \textrm{ outside } U(e) \\
     |id_{C,U(e),\epsilon}-1| & < & \epsilon \, \textrm{ for } x \in C,
   \end{array}
\end{equation}
which can be constructed as a sum of $\langle f_i,f_i \rangle_{red}$ with finitely many $f_i$ for any triple $(C,U(e),\epsilon)$. Given an operator $O$ of the full theory, one can induce a corresponding operator $O_{red}$ in the reduced theory by
\begin{equation}
  O_{red}:=\lim id_{C,U(e),\epsilon} O = \lim \sum_i \langle P(f_{i,C,U(e),\epsilon}),P(O^* f_{i,C,U(e),\epsilon}) \rangle_{red}.
\end{equation}

\subsection{Reducing Constrained Quantum Theories}

Let us briefly review the treatment of constraints in the construction of a reduced quantum theory. Given a self-adjoint anomaly free set $\mathcal C=\{C_i\}_{i \in \mathcal I}$ of constraint operators $C_i$, it is the objective of a gauge theory to construct the algebra Dirac observables, which are given by operators that commute with all constraints.

The canonical bilinear form for the Hilbert-$C^*$-module is built from two configuration variables $f_1,f_2$. Let us suppose that we have an approximate identity $id=\lim_{i \in \mathcal I} \sum_j \langle f^i_j,f^i_j \rangle_A$ for the algebra of Dirac observables, i.e. this approximate identity becomes a projector into the physical Hilbert space\footnote{This is of course only possible if the physical Hilbert space is a subspace of the kinematical Hilbert space.}. Having such a physical approximate identity, we can associate a Dirac-observable to each operator $O$ by:
\begin{equation}
  \lim_{i\in \mathcal I} id_i O = \lim_{i \in \mathcal I} \sum_j \langle f^i_j,f^i_j \rangle_A O = \lim_{i \in \mathcal I} \sum_j \langle f^i_j, o^* f^i_j \rangle_A.
\end{equation}
Using the correspondence $\langle P(f_1),P(f_2) \rangle_{red} : P(h) \mapsto P(\langle i(P(f_1)),i(P(f_2))\rangle_A i(h))$ between the bilinear structures of the unreduced theory and the reduced theory, we insert the approximate identity to induce the reduced Dirac observables using the bilinear structure $\langle .,. \rangle^o_{red}$, defined by
\begin{equation}
  \langle P(f_1),P(f_2) \rangle^o_{red} : P(h) \mapsto P(\lim_{i \in \mathcal I} id_i \langle i(P(f_1)),i(P(f_2))\rangle_A i(h)),
\end{equation}
by extending the span of $\langle .,.\rangle_{red}^o$ by linearity. Given a state $\Omega$ on the algebra of Dirac observables, we can use the induction
\begin{equation}
  \omega^o_{red}( \langle P(f),P(g) \rangle^o_{red}):=\Omega(\lim_{i \in \mathcal I} id_i \langle i(P(f)),i(P(g))\rangle_A)
\end{equation}
to construct the induced gauge invariant algebra. These induced states are then used for a GNS-conatruction to induce a representation.

If the operators $\langle .,. \rangle_A$ turn out to be true rank one operators in the full theory, then one can simplify this construction drastically. The observation that the operators $\langle f,g \rangle_A$ commute with all constraints, if $\pi_\omega(f)\Omega_\omega,\pi_\omega(g)\Omega$ are in the physical Hilbert space\footnote{$\Omega_\omega$ denotes the vacuum state of the cyclic representation $\Omega$ for a gauge invariant state $\omega$ and $\pi_\omega$ the according GNS-representation.} allows us to construct the Dirac observables of the full theory as a closure of the span of
\begin{equation}
  \langle f,g \rangle_A|_{f\Omega,g\Omega \in \mathcal H_{phys}}.
\end{equation}
Having a quantum embedding $(P,i)$, where $P$ is restricted to the configuration variables $f$ for which $f \Omega_\omega\in \mathcal H_{phys}$ and where $i$ takes values in these configuration variables then gives the usual induction for the algebra and representation. The restriction of the construction yields the reduced Dirac observables and their induced representation.

\subsection{A Combinatorial Quantum Theory}

Loop Quantum Gravity is a quantum theory that is built on a configuration space given by morphisms from the path groupoid in a base manifold to the gauge group of the Ashtekar connection. The overcountability of the set of piecewise analytic paths poses certain difficulties on the construction is this paper. In \cite{diffeo} we developed an idea based on observations in \cite{fairbairn} that allows for loop quantum gravity to be built from a combinatorial theory. The general framework of this class of quantum theories goes beyond the scope of this paper and will be published later, here we will only introduce the ingredients necessary for the construction in this paper.

\subsubsection*{Combinatorial Groupoid}

Let us consider the following groupoid: The unit set $\mathcal G^{(o)}$ is given by triples of integers $(N_1,N_1,N_3)$ denoted by $n$ for short. Consider the  finite sequences $(n_1,n_2,...,n_k)$ of triples of integers, which satisfy the pattern: given $n_i=(N_{i1},N_{i2},N_{i3})$ then $n_{i+1}$ is either of the form $((N_{i1}+/-1,N_{i2},N_{i3}))$ or $(N_{i1},N_{i2}+/-1,N_{i3})$ or $(N_{i1},N_{i2},N_{i3}+/-m)$, where $m$ is an integer being either $m=1$ or $0<m<N_{i3}$. let us define an equivalence relation $\sim$ between these sequences by
\begin{equation}
  \begin{array}{rcl}
    (n_1,n_2,...,n_{i-1},n_i,n_i,n_{i+1},...,n_k)&\sim&(n_1,n_2,...,n_{i-1},n_{i+1},...,n_k)\\
    (n_1,n_2,...,n_{i-1},n_i,n_{i+1},n_i,n_{i+2}...,n_k)&\sim&(n_1,n_2,...,n_{i-1},n_{i+2}...,n_k)
  \end{array}
\end{equation}
The groupoid set is then defined as the equivalence classes of sequences that satisfy the pattern rules. Clearly, there exists a shortest representative in each equivalence class, s.t. we can build a representation of the groupoid by denoting each equivalence class by its shortest representative. We will work in this representation. The source and range maps of the groupoid are given by
\begin{equation}
  \begin{array}{rl}
    s((n_1,n_2,...,n_k))&=n_1\\
    r((n_1,n_2,...,n_k))&=n_k,
  \end{array}
\end{equation}
the object inclusion map is given by
\begin{equation}
  e(n)=(n)
\end{equation}
and the composition is defined as concatenation and subsequent equivalenceing:
\begin{equation}
  (n_1,...,n_k) \circ (n_k,...,n_m)=(n_1,...,n_k,...,n_m)/\sim.
\end{equation}
The inverse can be determined to be:
\begin{equation}
  (n_1,...,n_k)^{-1}:=(n_k,n_{k-1},...,n_1).
\end{equation}
We call this groupoid {\bf combinatorial groupoid} throughout this paper.

\subsubsection*{Configuration Space}

Given a compact Lie-group $\mathbb G$, we can consider the morphisms of the combinatorial groupoid into this group. Since the combinatorial groupoid is generated by the set $\mathcal G$ of pairs of triples of integers that are of the form:
\begin{equation}
  \begin{array}{l}
    ((N_1,N_2,N_3),(N_1+1,N_2,N_3))\\
    ((N_1,N_2,N_3),(N_1,N_2+1,N_3))\\
    ((N_1,N_2,N_3),(N_2,N_2,N_3+m)) \,\textrm{ for }m \in \{1\}\cup \{n:2\le n \le N_3-1\},
  \end{array}
\end{equation}
we can write a groupoid morphism $A\in Hom(\mathcal G,\mathbb G)$ as a general map $A: \mathcal G \rightarrow \mathbb G$. The morphisms from any finite subset of $\mathcal G$ can be topologized by the product topology of $mathcal G$, which is compact. Since the finite subsets of $\mathcal G$ are partially ordered by the subset relation: $\gamma_1\le\gamma_2$ iff $\gamma_1\subset\gamma_2$, we can give $Hom(\mathcal G,\mathbb G)$ the structure of an inductive limit: $\lim_{\leftarrow \gamma} Hom(Gen(\gamma),\mathbb G)$\footnote{$Gen(\gamma)$ denotes the subgroupoid of $\mathcal G$ generated by the elements of $\gamma$.}. This allows us to give $Hom(\mathcal G,\mathbb G)$ a Tikhonow inductive limit topology, which is compact and Hausdorff because $\mathbb G$ is. This space $\mathbb X=Hom(\mathcal G,\mathbb G)$ together with the Tikhonow topology defines the configuration space.

A set $\gamma$ of finitely many generators $\gamma=(g_1,...,g_n)\subset \mathcal G$ together with a continuous complex valued function $f:\mathbb G^n \rightarrow \mathbb C$ defines a {\bf cylindrical function} by:
\begin{equation}
  Cyl=f(A(g_1),...,A(g_n)),
\end{equation}
which are always continuous. Moreover using the the sup-norm for functions on an arbitrary number $m$ of copies of $\mathbb G$ we can define a commutative $C^*$-algebra $\overline{Cyl}(\mathbb X)$ as the norm completion of the cylindrical functions. Using the techniques of \cite{ash-lew} one can verify that the $C^*$-algebra $\overline{Cyl}(\mathbb X)$ coincides with the algebra $C(\mathbb X)$ of continuous functions on the configuration space.

\subsubsection*{Finite Weighted Decomposition Functions}

A finite set $(g_1,...,g_n)$ of composable groupoid elements is called a decomposition of $g=g_1\circ g_2\circ ...\circ g_n$. We denote the set of all decompositions on the combinatorial groupoid by $Dec$. A map $d$ form the combinatorial groupoid into the decomposition set $Dec$ is called a {\bf decomposition function}, iff $g=dg_1\circ ... dg_n$ for all elements $g$ of the combinatorial groupoid.

Let us consider the weighted combinatorial groupoid, which is generated by triples $(t_1,g,t_2)$, where $t_1,t_2\in \mathbb R$ and $g$ is an element of the combinatorial groupoid; its unit set consists of pairs $(t,g)$ with $t\in \mathbb R$. The weighted source- and range maps are: $s(t_1,g,t_2)=(t_1,s(g))$ and $r(t_1,g,t_2)=(t_2,r(g))$. The composition law is given by $((t_1,g_1,s_1),...,(t_n,g_n,s_n))\circ((t_{n+1},g_{n+1},s_{n+1}),...,(t_m,g_m,s_m))=((t_1,g_1,s_1),...,(t_n,g_n,s_n),(t_{n+1},g_{n+1},s_{n+1}),...,(t_n,g_n,s_n))/\sim$, where we employ an analogous equivalence relation, that takes $(...,(t_1,g_1,-t_2),(t_2,g_2,t_3),...)\sim (...,(t_1,g_1\circ g_2,t_3),...)$ if $g_1,g_2$ are composable.

Given a real function $f$ on the unit set of the combinatorial groupoid $f:\mathcal G^{(o)} \rightarrow \mathbb R$ and a decomposition function $d$, we can define a {\bf weighted decomposition} $df$ on the weighted combinatorial groupoid by:
\begin{equation}
  \begin{array}{rcl}
  df & :&  (t_1,g,t_2) \mapsto ((t_1+f(s(dg_1)),dg_1,-f(r(dg_1))),\\
   && (f(s(dg_2)),dg_2,-f(r(dg_2))),...,(f(s(dg_n)),dg_n,t_2-f(r(dg_n))))/\sim.
  \end{array}
\end{equation}
The set of all finite weighted decompositions form a group. The morphisms $\theta(df)$ from the decomposition group to $Hom(\mathbb X)$ of the form
\begin{equation}
  \theta(df) A : g \mapsto \exp((t_1+f(s(dg_1)\xi_1) dg_1\exp(-f(r(dg_1))\xi_2)...\exp((t_2-f(r(dg_n)))\xi_k),
\end{equation}
where $\xi_i \in \mathfrak g$ defines a finite weighted decomposition function. These finite weighted decomposition functions form a group of homeomorphisms of $\mathbb X$, wich we denote by $\Theta$. $\Theta$ can be viewed as an inductive limit of decompositions that affect only groupoid elements whose expansion in the generators contain at least one element of a finite subset of the generators of the combinatorial groupoid. Due to the inductive limit and morphism structure into a compact group, one is able to give $\Theta$ a Tikhonow topology.

\subsubsection*{Canonical Quantum Algebra}

We have a commutative $C^*$-algebra $C(\mathbb X)$ of continuous functions on a compact Hausdorff space $\mathbb X$ and a compact group $\Theta$ of homeomorphisms of $\mathbb X$. This allows us to construct a crossed product $C^*$-algebra. We can construct cylindrical functions on $\mathbb X\times \Theta$ using the same technique as in the construction of cylindrical functions on $\mathbb X$. These are functions $f\in C(\mathbb G^n\times \mathbb G^m)$ of the form $f(A(g_1),...,A(g_n),P_1,...,P_m)$, where $P_i(g)$ is the elementary decomposition that splits precisely those elements of the combinatorial groupoid that pass through $s(g_i)$ into a part left $d_ig_l$ of $s(g_i)$ and a right part $d_ig_r$ and inserts a the group element: $P_i A (g)=A(d_ig_l)gA(d_ig_r)$. Moreover we are able to give these functions a convolution product by
\begin{equation}
  \begin{array}{rcl}
  f_1 * f_2 (A_1,...,A_n,P_1,...,P_m) &:=& \int d\mu_H(g_1) d\mu_H(g_m) f_1(A_1,...,A_n,g_1,...,g_n) \\ && P_1(g_1)...P_m(g_m) f_2(A_1,...,A_n,P_1,...,P_m).
  \end{array}
\end{equation}
Together with the natural involution and the norm defined by the supremum of the norms of all Hilbert space representations defines a natural noncommutative $C^*$-algebra $\mathfrak A(\mathbb X,\Theta)$ associated to the combinatorial groupoid.

$\mathfrak A(\mathbb X,\Theta)$ serves as the Weyl-algebra of the combinatorial system, where $C(\mathbb X)$ represents the configuration variables and the group $\Theta$ represents the the group of unitary Weyl transformations.

\subsubsection*{Unitary action of a Group}

There is an interesting group, whose action on $\mathfrak A(\mathbb X,\Theta)$ is ''unitary''. Consider a subgroup $\mathcal D$ of automorphisms of the combinatorial groupoid. One can easily define a unitary action $U$ of automorphisms $\phi \in \mathcal D$ on cylindrical functions by:
\begin{equation}
  \begin{array}{rl}
    \left(U^*(\phi)Cyl_{g_1,...,g_n}U(\phi)\right) (A)& := \, Cyl_{\phi(g_1),...,\phi(g_n)}(A)\\
    U^*(\phi) &:= \, U(\phi^{-1}).
  \end{array}
  \label{unit-act}
\end{equation}
This action can be extended by density to $C(\mathbb X)$. There is furthermore a natural extension of the action of $\mathcal D$ to $\Theta$ by ''translating'' the decomposition functions, i.e. $U^*(\phi)W(df)U(\phi):=W(\alpha(df))$, where $\alpha_\phi(df)$ decomposes a groupoid element $g$ into $d\phi(g)$ and inserts the weights $f(\phi(s(dg_i)))$. Since $C(\mathbb X)$ and $\Theta$ are the building blocks for $\mathfrak A(\mathbb X,\Theta)$, we can build elements of the form $f \circ U$, where $f\in \mathfrak A(\mathbb X,\Theta)$ and $U$ denotes the unitary action of an element of $\mathcal D$. The algebra spanned by the elements of this form is denoted by $\mathfrak A(\mathbb X,\Theta,\mathcal D)$.

Another unitary action of a groupoid can be achieved as follows: Consider an embedding $i$ of the combinatorial groupoid into a ''larger'' groupoid $\mathcal P$, s.t. the $i$-image of the combinatorial grouoid forms a subgroupoid of $\mathcal P$. Moreover consider a subgroup $\mathcal D^\prime$ of the automorphisms of $\mathcal P$. Let us define the space $\mathbb Y:=Hom(\mathcal P,\mathbb G)$ and construct a topology and cylindrical functions to obtain $Cyl(\mathbb Y)$ and $C(\mathbb Y)$, the analogues of $Cyl(\mathbb X)$ $C(\mathbb X)$ and let us topologize $\mathbb Y$ using the Tikhonow topology of the analogous projective limit. Let us use $i$ to embed $C(\mathbb X)$ into $C(\mathbb Y)$, which defines a unitary representation of $\Theta$ on $i(C(\mathbb X))\subset C(\mathbb Y)$. Let $\mathcal D^\prime$ define an action on $C(\mathbb Y)$ analogous to (\ref{unit-act}). This allows us to construct the natural extension of $\Theta$ to $\mathcal D(i(C(\mathbb X)))$ again by translating the action of $\Theta$ to $i(C(\mathbb X))$ and then by translating it back using $\phi^\prime,\phi^{\prime,-1}\in \mathcal D^\prime$. This defines an algebra $\mathfrak A(\mathbb X,i,\mathbb Y,\Theta,\mathcal D^\prime)$, which will be the structure that we use to construct Loop Quantum Gravity. We call the algebra $\mathfrak A(\mathbb X,i,\mathbb Y,\Theta,\mathcal D^\prime)$ spanned by the embedding $i$ and the action $\mathcal D^\prime$ out of the algebra $\mathfrak A(\mathbb X,\Theta)$

\subsubsection*{Canonical Representation}

The canonical uniform measure $\mu_o$ on $\mathbb X$ is defined through the positive\footnote{The functional maps positive elements of $Cyl$ into $\mathbb R^+$.} linear functional on that we define on the space of cylindrical functions as:
\begin{equation}
  \omega(Cyl):=\int d\mu_H(h_1)...d\mu_H(h_n) Cyl_{g_1,...,g_n}(h_1,...,h_n), \label{omega-o}
\end{equation}
and which we extend by density to $C(\mathbb X)$. The Schr\"odinger type representation of $\mathfrak A(\mathbb X,\Theta)$ is defined through the following inner product on the pre-Hilbert space $Cyl(\mathbb X)$:
\begin{equation}
  \langle Cyl_1, Cyl_2 \rangle := \omega(\overline{Cyl_1}Cyl_2),
\end{equation}
which allows us to construct a Hilbert-space $\mathcal H=L^2(\mathbb X,\mu_o)$ as the completion of $Cyl(\mathbb X)$ in the inner product norm. The canonical representation $\pi_o$ of $\mathfrak A(\mathbb X,\Theta)$ on $\mathcal H$ is then defined through the covariant pair of representations of $Cyl(\mathbb X)$ and $\Theta$:
\begin{equation}
  \begin{array}{rl}
    \pi_o(Cyl_1) Cyl_2 &:=\, Cyl_1 Cyl_2\\
    \pi_o(W(df)) Cyl   &:= \theta^*(df) Cyl,
  \end{array}
\end{equation}
where the extension by density of $Cyl(\mathbb X)$ in $C(\mathbb X)$ as well as $\mathcal H$ is used. $\pi_o$ clearly defines a unitary action of $\Theta$ on $\mathbb X$.

The canonical representation also denoted by $\pi_o$ of $\mathcal D$ on $\mathcal H$ defined through
\begin{equation}
  \pi_o(U(\phi)) Cyl_{g_1,...,g_n} := Cyl_{\phi(g_1),...,\phi(g_n)}
\end{equation}
is clearly unitary due to the invariance of $\mu_o$ under automorphisms of the combinatorial groupoid.

The analogous construction defines a canonical representation of $\mathfrak A(\mathbb X,i,\mathbb Y,\Theta,\mathcal D^\prime)$ on $L^2(\mathbb Y,d\mu_Y)$, where $d\mu$ is defined through a functional in complete analogy to the definition of $d\mu_o$ in (\ref{omega-o}). Notice that the ground state $Cyl=1\in \mathcal H$ is invariant under $\mathcal D^\prime$.

\subsection{Combinatorial Approach to Loop Quantum Gravity}

The similarity between Loop Quantum Gravity and the combinatorial theory that we described in the previous section is rather obvious. Let us now construct Loop Quantum Gravity as an algebra of the kind $\mathfrak A(\mathbb X,i,\mathbb Y,\Theta,\mathcal D^\prime)$. The gauge group of Loop Quantum Gravity is $SU(2)$ such that $\mathbb X$ is specified as the set of homeomorphisms from the combinatorial groupoid to $SU(2)$.

We construct an explicit embedding of the combinatorial groupoid into the groupoid of piecewise analytic paths $\mathcal P$ in section \ref{scaffold} for the case that the base manifold $\sigma=\mathbb R^3$. Using this embedding $i$, we can construct the configuration space of Loop Quantum Gravity with $\mathbb Y=Hom(\mathcal P,SU(2))$.

Let us now consider the following groupoid: The groupoid set consists of all finite collections of analytical surfaces (which we assume to be homeomorphic to a disc) $(S_1,...,S_m)$ that are analytically embedded into $\sigma$ together with finite collections of analytical paths $(e_1,...,e_n)$, which are also analytically embedded into $\sigma$, i.e. collections
\begin{equation}
  \mathcal G^{(o)}=\{(e_1,...,e_n,S_1,...,S_m):\,e_i\subset \sigma, \, S_j\subset \sigma\}.
\end{equation}
The groupoid then consists of all pairs of these collections $((e_1,...,e_n,S_1,...,S_m),(e^\prime_1,...,e^\prime_n,S^\prime_1,...,S^\prime_m))$ for which there exists a homeomorphism of $\sigma$ that maps $(e_1,...,e_n,S_1,...,S_m)$ into $(e^\prime_1,...,e^\prime_n,S^\prime_1,...,S^\prime_m)$. The source- resp. range maps return the first resp. second collection out of these pairs. The composition law is
\begin{equation}
  \begin{array}{l}
   ((e_1,...,S_1,...),(e_2,...,S_2,...))\circ((e_2,...,S_2,...),(e_3,...,S_3,...)):=\\
   ((e_1,...,S_1,...),(e_3,...,S_3,...)).
 \end{array}
\end{equation}
We denote this groupoid by $\mathcal D^\prime$. One can span an algebra using an embedding and a groupoid as well as using a group. Using the embedding $i$ from section \ref{scaffold} of the combinatorial theory $\mathfrak A(\mathbb X,\Theta)$, we can use the canonical action of the groupoid as pairs consisting of an automorphism of the path groupoid together with an map of Weyl-operators to span the quantum algebra of Loop Quantum Gravity as $\mathfrak A(\mathbb X,i,\mathbb Y,\Theta,\mathcal D^\prime)$.

The canonical representation $\pi_o$ defines the canonical representation of an algebra underlying Loop Quantum Gravity (for the description of this algebra see section \ref{obs-alg}), whose vacuum vector $Cyl=1\in \mathcal H$ is obviously invariant under $\mathcal D^\prime$.

\section{Reduction Maps for Cosmology}

We will construct general reduction maps for LQG in this section, which we apply in the next section to extract a cosmological sector.

\subsection{Introductionary Considerations}\label{intro}

The cosmological sector of General Relativity arises as the sector that is invariant under a group of spatial symmetries. These spatial symmetries however are just a subgroup of the diffeomorphism group, which is part of the gauge group\footnote{Following the treatment of constraints by Dirac, we call the entire group that is generated by the Gauss-, diffeomorphism- and scalar constraint gauge group of General Relativity in Ashtekar variables. Since all of these constraints generate unobservable transformations, it is legitimate to call all these transformations gauge transformations.} of General Relativity. Thus it seems at first sight spurious to assume a symmetry that can be viewed as a subgroup of a gauge group, since we are interested in a quantum theory, which is built from gauge invariant quantities and these quantities are already necessarily invariant under the spatial symmetry group.

What we seek to construct is however not a model that is built from the heuristics about these symmetric states, but we want to preform the quantum analogue to to the phase space reduction that results classically when the spatial symmetries are imposed. This can be stated with the catchy phrase: ''We want to construct the diffeomorphism invariant version of this symmetry reduction.''

These classical symmetry reduced models have a finite dimensional configuration space, thus we want to construct a reduced quantum algebra from the quantum algebra of the full theory of Loop Quantum Gravity that can be viewed as a quantization of the reduced phase space. We outlined a strategy for this procedure in \cite{reduction}, which we interpreted physically as a reduction of the sensitivity of the measurements at our disposal to observables on the reduced phase space. For the purposes here, we have to change this interpretation, because the solutions to the diffeomorphism constraint are distributional and can as such not be evaluated at individual points of the phase space, but only over open sets.

The distributionality of the solutions to the diffeomorphism comes about, because these functions are constructed as sums over all diffeomorphisms acting on cylindrical functions\footnote{This statement is imprecise, because to carry the group averaging out one has to decompose the diffeomorphism group into automorphisms of the graph that underlies the particular cylindrical function and the quotient under these automorphisms. The two factors are then treated differently under the rules of group averaging.}. However, we can also proceed differently and (partially) gauge fix the diffeomorphisms. This is to say, that we assign exactly one representative cylindrical function to each gauge orbit of cylindrical functions. Since the gauge orbits of a cylindrical function are in a one to one correspondence to the distributions that we obtain by group averaging the diffeomorphisms, we have to assign to each of the group averaged quantities exactly one cylindrical function whose group averaging yields precisely this solution.

Most differences of this approach compared with standard LQC are due to the fixing of the diffeomorphisms. Each fixing of the diffeomorphisms results in different rule for imposing the symmetry reduction of the noncommutative phase space. The different symmetry reduction yield in general different results. All of these are reductions of the phase space, the interpretation which physical sector this reduction corresponds to has to be determined by measurements.

What does this gauge fixing mean for measurements? An observer has certain measurements at his disposal and having a gauge theory means that even after preforming a complete set of measurements with these observables, he can still not solve unambigously for the degrees of freedom in the underlaying theory. Fixing a gauge means providing a set of relations such that a complete set of measurements together with these relations can be solved for the theoretical degrees of freedom, which are then in this particular gauge. This means for our reduction that we have to provide a gauge in which we prefrom the phase space reduction. Then, by construction, there will be an observer and a gauge such that we are resolving exactly the reduced phase space.

In \cite{reduction} we constructed the quantum reduction map from equivalence classes of observables on the configuration space, which differ only by their dependence on the complement of the reduced configuration space. However, the evaluation of an observable on a homogeneous connection depends obviously on the gauge that we choose for the diffeomorphisms. Let us illustrate this on a simple example: Consider a chart in which the $\omega^i = dx^i$. Let us consider a graph $\gamma$ around a unit square in the $x_1,x_2$-plane in this chart\footnote{We denote a straight edge in a chart by its initial and final point, i.e. in a chart $(U,\varphi)$ the expression $e=((i_1,i_2,i_3),(f_1,f_2,f_3))$ is a shorthand for $e=\{\varphi(i^a+t(f^a-i_a))\in U: 0 \le t \le 1 \}$}:
$$
  \begin{array}{l}
    \gamma=(e_1,...,e_4)=\\
    (((0,0,0),(1,0,0)),((1,0,0),(1,1,0)),((1,1,0),(0,1,0)),((0,1,0),(0,0,0)))
  \end{array}
$$
as well as the graph $\gamma^\prime=(f_1,...,f_4)$ around the structure shifted by one unit in $x_3$-direction:
$$
  \begin{array}{l}
    \gamma^\prime=(f_1,...,f_4)=\\
    (((0,0,1),(1,0,1)),((1,0,1),(1,1,1)),((1,1,1),(0,1,1)),((0,1,1),(0,0,1))).
  \end{array}
$$
Then any cylindrical function $T_\gamma(A)=f(h_{e_1}(A),...,h_{e_4}(A))$ on $\gamma$ will coincide with the cylindrical function $T_{\gamma^\prime}(A)=f(h_{f_1}(A),...,h_{f_4}(A))$. However, if we apply a diffeomorphism $\phi$ that acts in this chart $(U,\varphi)$ as
$$
  \phi: \varphi(x_1,x_2,x_3) \mapsto \varphi(x_1(1+x_3^2),x_2(1+x_3^2),x_3)
$$
that leaves $\gamma$ invariant but stretches the edges of $\gamma^\prime$ by a factor of two, then $\alpha_\phi(T_\gamma(A))$ and $\alpha_\phi(T_{\gamma^\prime}(A))$ will not coincide on homogeneous connections for a general $f$\footnote{Here $\alpha_{\phi}$ denotes the action of a diffeomorphism on a cylindrical function by $\alpha_{\phi}(Cyl_\gamma):=Cyl_{\phi(\gamma)}$}, although $T_{\gamma}(A)$ and $T_{\gamma^\prime}(A)$ do coincide on homogeneous connections. Obviously, we can also turn this argument around and start with two cylindrical functions $\alpha_\phi(T_\gamma(A))$ and $\alpha_\phi(T_{\gamma\prime}(A))$ that do not coincide on homogeneous connections and apply the inverse $\phi^{-1}$ of $\phi$ to obtain two cylindrical functions that coincide on homogeneous connections.

Our proposed solution to this problem stems from the observation that an observer has only cylindrical functions on knot classes as measurements at his disposal and not particular graphs. So if an observer is asked about determining the homogeneous part of a connection, then needs to know about the gauge fixing of the diffeomorphisms, i.e. he needs to be provided with a particular embedded representative graph for each knot class. However, given this gauge he can easily determine whether the system is in a homogeneous state or not, by simply testing whether the relations among the observables that are implied by homogeneity and the gauge are satisfied in his measurements or not. These relations for connection observables are however not necessarily relations for geometrical quantities. This can be seen by the fact that we can choose a gauge fixing for the diffeomorphisms such that a given region $R_1 \subset U$ is avoided by the embedding of graphs in this gauge and another region $R_2\subset U$ is densely populated with edges. Then the expectation value of any area or volume in $R_1$ will necessarily vanish, while the expectation value of any such area in $R_2$ will be ''large''. Thus the relations for geometrical observables implied by homogenuity do not hold. We will however demand that our gauge fixing for the diffeomorphisms is such that the relations implied by homogenuity are satisfied at least approximately at a certain scale.

The specification of a region or a surface through an embedding is of course not background independent. Physically a region or surface is specified by matter residing thereon, e.g. a region may be specified by all the vertices in a graph, where the field strength of a scalar field takes certain values and a surface may be specified as the edges that link a vertex inside with a vertex outside, being the analogue to the boundary of a region specified by the occupied vertices. Within the framework of this work, it turns out to be much less complicated to impose a ''superficial homogeneity'' defined by an approximate homogeneity of embeddings. The connection to the physical notion is as follows: The regions in which a scalar field takes values in a certain range then defines under a certain coarse graining an embedded region. The ''superficial homogeneity'' implies physical homogeneity for certain gauge fixings of the diffeomorphisms. This connection between the two notions of homogeneity is the reason why we impose the simpler one for the construction of the gauge fixing for the diffeomorphisms.

\subsection{Strategy}

Given a particular gauge fixing for the diffeomorphisms, we can apply the techniques of \cite{reduction} to construct a cosmological sector of Loop Quantum Gravity. We just saw that the choice of gauge has consequences for the geometrical observables, at least at the level of superficial homogeneity, which we will be the notion that we consider here.

Thus, our first step consists of constructing a gauge fixing of the diffeomorphisms such that the geometrical observables become approximately homogeneous in a certain region under a particular class of coarse grainings. The idea is to construct a scaffold of allowed vertices and links among them, such that any knot class of graphs can be embedded into a finite region. For this purpose we will start out with a homogeneous chart\footnote{A in the previous section, we call a chart homogeneous, if the one forms $\omega^i$, that define homogeneity, take the special form $\omega^i=dx^i$.} $U,\phi$ and construct a regular cubic lattice therein. This lattice has only six-valent vertices and is thus not able to accommodate for the embeddings of graphs with higher valent vertices. Thus, we have to add additional links between the vertices of the regular lattice, such that for any valence $n$ and any number $m$ of vertices there exists a finite region such that this region contains at least $m$ vertices which have at least valence $n$.\footnote{This does not yet ensure that any graph with $m$ vertices of at most valence $n$ can be embedded, because we did not yet consider the knotting of the graphs, however we will take care of this issue in our construction.} Once we have this scaffold that allows to embed any graph into a finite region, we can define a map $R(\gamma)$ that assigns to each graph a region that it can be embedded into. Using these two ingredients, we will be able to define a quantum symmetry reduction $P$ by assigning to each cylindrical function depending on a graph the average of this cylindrical function under all embeddings of the knot class of this graph\footnote{We will refer to graphs as {\bf tame} if all their edges and vertices are part of the scaffold.} into the region $R(\gamma)$.\footnote{The choice of $R$ should obviously be such that any finite region in $\mathbb R^3$ is covered by some large enough graph, such that the superficial homogeneity is satisfied.} The map $i$ will then be constructed as a linear map, that assigns each function of the reduced connection that is in the image if $P$ exactly one representative spin-network function on a graph that is embedded into the scaffold and whose dependence on the homogeneous connection is precisely that of the original function.

Let us summarize the steps that we will work through in the next two subsections. These steps are not particularly geared to cosmological models, but slight modifications can be used for more general extraction procedures of mini-superspaces from Loop Quantum Gravity:

\begin{enumerate}

  \item We construct a scaffold, which is an infinite collection of embedded vertices and links embedded in the base manifold. This scaffold has to be large enough so there exists a function $R(\gamma)$ that assigns each graph $\gamma$ a region such that the scaffold restricted to this region is larger than $\gamma$, i.e. $\gamma$ can be embedded into this part of the scaffold. While this procedure is sufficient for a noncompact base manifold, one has to reverse it in the compact case in order to avoid accumulation points of edges and vertices of the scaffold. This can be achieved by defining a size $S:$graphs$\rightarrow\mathbb N$ of a graph\footnote{Using e.g. a particular procedure to embed a graph into the scaffold results in a minimal cube in the regular lattice, whose edge length can be taken as such a number $S$.} and defining a family of scaffolds such that each graph of size $n$ can be embedded into the $n$th scaffold.

  \item We construct $P$ by assigning to each cylindrical function $Cyl_\gamma$ the restriction of this function to homogeneous connections. This dependence can in general not be calculated explicitly, because holonomies along arbitrary curves can not be evaluated even when the connection is homogeneous. This is the reason, why we have to gauge-fix the diffeomorphisms to a scaffold such that all holonomies can be computed explicitly for homogeneous connections. Thus we define $P$ in this gauge, which means that we may have to apply a diffeomorphism $\phi$ such that $\phi(\gamma)$ lies in the scaffold.

  \item The construction of $P$ is obviously linear, since the restriction of a function to a part of its domain is a linear operation. However, we have to construct a second linear map $i$ that assigns each restricted function in the image of $P$ exactly one representative such that $P\circ i=id_{img(P)}$ is satisfied. This can be achieved by taking a linearly independent set of functions, that spans the image of $P$. Then using the axiom of choice, there exists a map $i$ from this linearly independent to the domain of $P$ such that $P \circ i = id$ is satisfied on this set. Then $i$ is defines as the linear extension of this map to the entire image of $P$.

\end{enumerate}

This construction yields a quantum reduction map given by a pair $(P,i)$ that we then use to reduce Loop Quantum Gravity to a cosmological sector and induce its Hilbert space representation. It is obvious that the homogeneity assumption is not background-independent and thus different gauge fixings for the diffeomorphisms will in general yield different cosmological sectors.

\subsection{Construction of the Scaffold\label{scaffold}}

We assumed a Bianchi I model, such that the base manifold $\Sigma$ is $\mathbb R^3$. Let us fix a global chart $U,\varphi$ such that $U=\mathbb R^3\sim\Sigma$ and that is homogeneous, i.e. $\varphi^* \omega^i = dx^i$ for the forms $\omega^i$ that define homogeneity. We will later assume that the connection components are homogeneous w.r.t. these $\omega^i$, i.e. the components of the connection are linear combinations of the $\omega^i$.

Let us use our shorthand for the notation of straight curves, i.e. we make the identification:
\begin{equation}
  ((i_1,i_2,i_3),(f_1,f_2,f_3)):=\{\varphi(i^a+t(f^a-i^a))\in\Sigma:0 \le t \le 1\}.
\end{equation}
Our first step in the construction of the scaffold is the construction of a regular lattice of fiducial length $l_o$\footnote{We endow $U$ with an unphysical Euklidean metric $\delta_{ij}dx^i\vee dx^j$.}. This consists of the three families of edges:
\begin{equation}
  \begin{array}{rl}
    e^1_{abc}:=& ((l_o a, l_o b, l_o c),(l_o(a+1),l_o b, l_o c))\\
    e^2_{abc}:=& ((l_o a, l_o b, l_o c),(l_o,l_o (b+1), l_o c))\\
    e^3_{abc}:=& ((l_o a, l_o b, l_o c),(l_oa,l_o b, l_o (c+1))),
  \end{array}
\end{equation}
where $a,b,c \in \mathbb Z$ as well as the family of vertices
\begin{equation}
  v_{abc}:=\varphi(l_o a,l_o b,l_o c),
\end{equation}
where again $a,b,c \in \mathbb Z$.

So far, we have only constructed a regular lattice with six-valent vertices, but in order to embed the graphs from Loop Quantum Gravity, we need to be able to embed graphs with vertices of arbitrary valence. This is done by adding a family of extra links:
$$
  l_{abcn}
$$
between $v_{a,b,c}$ and $v_{a+n,b,c}$ where $a,b,c,n \in \mathbb Z$ and $n < a$, that do neither intersect with each other nor intersect with the regular lattice. A particular choice can be constructed as follows:
\begin{enumerate}

  \item For each positive integer $n$ consider a ''bridge'' of length $l_o n$
        $$ b_{abcn} := e^2_{abc} \cup ((l_o a,l_o(b+1),l_o c),(l_o(a+n),l_o(b+1,l_o c))) \cup e^2.$$

  \item Rotate this bridge by an angle $\alpha_{an}:=\frac\pi 2 \frac{n}{a^2}$ around the translation of the $x_1$-axis into $x_2=b,x_3=c$. Clearly all $\alpha_{an}$ are distinct, because $n < a$ by assumption. Thus, the extra links do not intersect. Moreover, since $0 < \alpha < \frac \pi 2$, the rotated bridges do not intersect with the lattice.

  \item The resulting rotated bridges are then defined to be our extra links:

\end{enumerate}

\begin{equation}
    \begin{array}{l}
      l_{abcn}= \\
      ((l_o a,l_o (b+\cos(\alpha_{a,n})),l_o (c+\sin(\alpha_{a,n}))),((l_o a,l_o (b+1+\cos(\alpha_{a,n})),l_o (c+\sin(\alpha_{a,n})))))\\
       \cup ((l_o a,l_o(b+1),l_o c),(l_o(a+n),l_o(b+1,l_o c)))\\
       \cup ((l_o (a+n),l_o (b+\cos(\alpha_{a,n})),l_o (c+\sin(\alpha_{a,n}))),((l_o (a+n),l_o (b+1+\cos(\alpha_{a,n})),l_o (c+\sin(\alpha_{a,n}))))).
    \end{array}
\end{equation}

Now we have all the ingredients to define our scaffold:
\begin{defi}
  $\bullet$ The {\bf scaffold} consists of the set of vertices $V=\{v_{abc}:a,b,c\in \mathbb Z\}$ as well of the set of edges $E=\{e^i_{abc}:i=1,2,3;a,b,c \in \mathbb Z\}\cup\{f_{abcn}:a,b,c\in \mathbb Z; n \in \mathbb N; n < a\}$
\end{defi}
We notice (1) that the scaffold is not a graph, because it contains an infinite number of edges and (2) that the scaffold, although containing vertices of arbitrary valence, does it does not contain an accumulation point of edges.

Let us now verify that any knot class can be embedded into this scaffold. Given the knot class of a graph, particularly given a projection of a graph $\gamma$, let us consider the following procedure:
\begin{enumerate}

  \item Label the vertices of $\gamma$ by natural numbers $1,...,N$, i.e. $V_\gamma=\{v_1,...,v_N\}$; label the edges of $\gamma$ by natural numbers $1,...,M$, i.e. $E_\gamma=\{e_1,...,e_M\}$.

  \item Let $K=N+2 M$ be the number of vertices plus twice the number of edges. Then embed the vertices by $i: v_n \mapsto v_{K+n,0,0}$.

  \item Each edge $e_a\in E_\gamma$ is split into three parts $e_{a,i},e_{a,m},e_{a,f}$ which are connected by vertices $v_{a,i}$ between $e_{a,i}$ and $e_{a,m}$ and $v_{a,f}$ between $e_{a,m}$ and $e_{a,f}$. This splitting can be chosen such that there are no overpasses of any pieces over the $e_{a,i}$ and $a_{a,f}$ in the given projection of $\gamma$, meaning conversely that all the overpasses are among the $e_{a,m}$.

  \item Extend $i$ such that the sets $v_{a,i}$ and $v_{a,i}$ of additional vertices are embedded into the scaffold-vertices $v_{K+N+1,0,0},...,v_{2K,0,0}$. Furthermore extend $i$ to the sets $e_{a,i}$ and $e_{a,f}$ by assigning the respective link $l_{defm}$ that connects the embedded boundary vertices.

  \item Define the parallel projection $P:(x,y,z)\mapsto (x,y)$ that assigns an overpass of $(x_1,y_1,z_2)$ over $(x_2,y_2,z_2)$ if $z_1 > z_2$. Take the image of the objects for which $i$ is defined so far and project them using $P$. This generally results in a nontrivial projection containing a number of overpasses of the $e_{a,i},e_{a,f}$. We notice that the projection does not contain over passings except trivial ones at the vertices $i(v_1),..,i(v_n)$ for $x_1<l_o(K+N+1/2)$, we can view the part of the projection for $x_1>l_o(K+N+1/2)$ as a braid $B_1$.

  \item Remove the vertices $v_a$ as well as the edge pieces $e_{a,i}$ and $e_{a,f}$ from the given projection of $\gamma$, but keep the $v_{a,i},v_{a,f}$ fixed. This defines a braid $B_2$.

  \item We notice that any braid with fixed boundaries can be embedded into a regular cubical lattice. Thus, attach such an embedding of the inverse braid of $B_1$ to the image of $i$ and then attach a the braiding $B_2$ onto these. This extends the embedding $i$ to the set $e_{a,m}$ of ''middle pieces of the edges in $\gamma$.

\end{enumerate}

Having this procedure at our disposal, we can construct an embedding $i$ for any given knot class of a graph into the scaffold. However, this procedure does in general not yield the simplest possible embedding $i$, however this is not necessary for our purposes. Notice that the edges in the scaffold are oriented, i.e. $((i_1,i_2,i_3),(f_1,f_2,f_3))$ is oriented from $i$ to $f$, which is important for the calculation of holonomies.

\subsubsection{Scaffold Observable Algebra}
\label{obs-alg}

Let us fix the observable algebra on the scaffold, that we want to subject to our symmetry reduction procedure. The heuristic idea is to consider a holonomy-flux-Weyl-algebra, where the holonomies are scaffold holonomies and the Weyl-operators are exponentials of fluxes on umbrella shaped regions: Given a vertex in the scaffold, we call a piecewise analytic surface ''umbrella shaped'' if it intersects precisely one adjacent edge transversally. The orientation of these surfaces is choosen such that the Weyl operators act as left $SU(2)$- translations on precisely one the holonomy along the transversally intersecting edge.

There are many ways to construct an acceptable $C^*$-algebra for quantum field theories, which are generally inequivalent as is known by Haag's theorem for background dependent quantum field theories: Using the Schr\"odinger representation of the Weyl-system of a free Klein-Gordon theory one can work in close analogy to Loop Quantum Gravity:

The fundamental configuration variables are modes $\phi(f):=\int_\Sigma d^3\sigma f(\sigma) \phi(\sigma)$, where the modes $f$ satisfy certain fall-off conditions. A cylindrical function $Cyl$ is a functional of the field $\phi$, that has the same dependence on $\phi$ as $F(\phi(f_{i_1}),...,\phi(f_{i_n}))$, where $F: \mathbb R^n\rightarrow \mathbb C$ is continuous (and grows less than exponentially) and $n \in \mathbb N_o$. It is often useful to use a narrower definition by restricting the set of modes to the eigenfunctions of a one-particle Hamiltonian that satisfy the fall-off conditions, so there is a countable set of modes $\{f_i\}_{i=1}^\infty$, which is complete and orthonormal in the one-particle inner product $(f_i,f_j)=\delta^{Kron.}_{ij}$. The finite sets $\{\phi(f_{i_1}),...,\phi(f_{i_n})\}$ are the analogues of graphs, which are partially ordered due to the subset relation. The elementary Weyl-operators are the exponentials $w_i(\mu):=\exp(i \mu\pi(f_i))$ of the conjugate momenta $\pi$, where $\pi(f_i)=\int_\Sigma d^3\sigma f_i(\sigma) \pi(\sigma)$, which are supposed to be unitary and to satisfy the Weyl-commutation relations: $w_i(\mu) F(...,\phi(f_i),...) w^*_i(\mu) = F(...,\phi(f_i)-\mu,...)$. Using the ''unitarity'' of the Weyl-operators $w_i^*(\mu)=w_i(-\mu)=w_i(\mu)^{-1}$, we find that finite sums of cylindrical functions and Weyl-operators $\sum_{i=1}^k Cyl_i w_i$ are closed under multiplication and that $F(\phi(f_{i_1}),...,\phi(f_{i_n}))^*:=\overline{F(\phi(f_{i_1}),...,\phi(f_{i_n}))}$ and $w_i^*(\mu)=w_i(-\mu)$ defines an involution. Given a positive real number $a_i$ for each mode $f_i$, one can define a Gaussian vacuum state:
$$
  \begin{array}{rcl}
    \omega(\sum_{i=1}^k Cyl_i w_i)&:=&\sum_{i=1}^k \int N(a_{i_1})dx_{i_1} e^{-\frac{a_{i_1}}{2} x_{i_1}^2}...N(a_{i_1})dx_{i_1} e^{-\frac{a_{i_1}}{2} x_{i_1}^2}\\
     && F(x_{i_1},...,x_{i_n}) e^{-\frac{a_{i_1}}{2} (x_{i_1}-\mu_{i_1})^2}...e^{-\frac{a_{i_n}}{2} (x_{i_n}-\mu_{i_n})^2}.\\
     &=:&\langle\Omega_\omega, \pi_\omega(\sum_{i=1}^k Cyl_i w_i) \Omega_\omega\rangle
  \end{array}
$$
The Hilbert space $\mathcal H_\omega$ constructed of the finite sums $\sum_{i=1}^k Cyl_i w_i$ and this vacuum state has the GNS representation, which is spanned by cylindrical functions, i.e. $\pi_\omega(Cyl) \Omega_\omega$ turns out to be dense in $\mathcal H_\omega$. Moreover, using the rank-one operators $|\pi_\omega(Cyl_1)\Omega_\omega\rangle\langle \pi_\omega(Cyl_2) \Omega_\omega|$, we can give the cylindrical functions the structure of a pre-Hilbert-pre-$C^*$-module in the obvious way by setting $\langle Cyl_1,Cyl_2\rangle_{\mathfrak A}:=|\pi_\omega(Cyl_1)\Omega_\omega\rangle\langle \pi_\omega(Cyl_2) \Omega_\omega|$ and using the action of $\pi_\omega$ thereon.

This transformation group structure allows us to define Rieffel's\cite{rieffel}  approximate identity $id_{\epsilon,C,U(1)}$, indexed by a tolerance $\epsilon >0$, compact sets $C$ of the locally compact configuration space and open neighborhoods $U(1)$ of the unit element of the Weyl-group on each graph as $id_{\epsilon,C,U(1)} = sum_{j=1} \langle Cyl_1,Cyl_2\rangle_{\mathfrak A}$. Note that the configuration space can be approximated by increasing compact sets, because it is a compact space itself due to the fact that $Cyl: \phi \mapsto 1$ is a cylindrical function on every graph. This allows us to construct an approximate identity $id_{\gamma,\epsilon,C_\gamma,U_\gamma(1)}$ by considering Rieffel's approximate identity for each graph. This allows us to define the observable $C^*$-algebra as the operator norm completion of $\pi_\omega(\sum_{i=1}^n Cyl_i w_i) id_{\gamma,\epsilon,C_\gamma,U_\gamma(1)}$ in $\mathcal B(\mathcal H_\omega)$. Using the pre-Hilbert-pre-$C^*$-module structure on the cylindrical functions one can establish a strong Morita equivalence with $\mathbb C$, which reflects Segals theorem \cite{segal} of the uniqueness of the representations of the infinite-dimensional CCR once a dynamics that factorizes over the set of modes is chosen, because the domain of one particle Hamiltonian is encoded in the choice of modes $\{f_i\}_{i=1}^\infty$ and its action is (partially) encoded in the positive real numbers $\{a_i\}_{i=1}^\infty$.

This construction can now be generalized to give a precise definition of the scaffold algebra by replacing the countably infinite set of modes with the countably infinite set of edges in the scaffold and the group $\mathbb R$, the range of the mode observables, by the gauge group $SU(2)$:

The elementary configuration observables on the scaffold are the (matrix elements of the) holonomies along edges in the scaffold, a graph is a finite set of edges $\{e_{i_k}\}_{k=1}^n$ in the scaffold and a cylindrical function $Cyl$ is a functional of the scaffold holonomies that has the same dependence on the scaffold holonomies as $F(h_{e_{i_1}},...,h_{e_{i_n}})$, where $F:SU(2)^n \rightarrow \mathbb C$ is continuous and $n \in \mathbb N_o$. The fundamental Weyl-operators act as left $SU(2)$-translations on the holonomies $w_i(g) h_{e_i} w_i(g)^*=g h_{e_i}$. The elementary observable algebra is $\sum_{i=1}^n Cyl_i w_i$, where the $w_i$ furnish a ''unitary '' representation of $SU(2)$, i.e. $w_i(g)^*=w_i(g^{-1})=w_i(g)^{-1}$, and the involution is $F(h_{e_{i_1}},...,h_{e_{i_n}})^* = \overline{F(h_{e_{i_1}},...,h_{e_{i_n}})}$. The canonical state on this algebra is:
$$
  \omega(\sum_{i=1}^n Cyl_i w_i):=\sum_{i=1}^n \int d\mu_H(g_1)...d\mu_h(g_{i_n}) Cyl_i (g_1,...,g_{i_n}),
$$
where $d\mu_H(g)$ denotes the Haar measure of $SU(2)$. $\omega$ leads to the canonical representation $\pi_o$ in which the spin networks $SNF$ form a dense orthonormal set of vectors $\pi_o(SNF)$ in the scaffold Hilbert space $\mathcal H_o$, whose inner product turns out to be $\langle \pi(Cyl_1) \Omega_o,\pi(Cyl_2)\Omega_o\rangle = \int d\mu_H(g_1)...d\mu_H(g_n) \overline{Cyl_1(g_1,...,g_n)}Cyl_2(g_1,...,g_n)$. This allows the construction of a pre-Hilbert-pre$C^*$-module for the sums $\sum_{i=1}^n Cyl_i w_i$ given by the cylindrical functions by setting $\langle Cyl_1,Cyl_2\rangle_{\mathfrak A}:=|\pi_o(Cyl_1)\Omega_o\rangle\langle \pi_o(Cyl_2) \Omega_o|$ and using the canonical representation $\pi_o$ thereon. We are thus able to apply Rieffel's construction of an approximate identity for each graph $\gamma$ and hence find an approximate identity $id_{\gamma,\epsilon,C_\gamma,U_\gamma(1)}$ for the entire observable algebra. We are thus able to define the $C^*$-algebra for the scaffold as the operator norm completion of the elements of the form $\sum_{i=1}^n Cyl_i w_i id_{\gamma,\epsilon,C_\gamma,U_\gamma(1)}$ in $\mathcal B(\mathcal H_o)$. The strong Morita equivalence between the scaffold algebra and $\mathbb C$ that is inferred by the cylindrical functions provides a uniqueness theorem for the representation of the scaffold algebra.

Let us now consider the relation between the scaffold algebra defined here and the observable algebra that underlies loop quantum gravity: The holonomy-flux-Weyl-algebra is constructed by smearing the electric fields on open piecewise analytic 2-dimensional surfaces. However, given any open piecewise analytic curve $c$ there is a Weyl-operator that corresponds to an electric field smeared on this 1-dimensional quasi-surface: For any $c$ there exists an open piecewise analytic surface $S$ such that every interior point of $c$ is also an interior point of $S$ and the boundary points of $c$ are in the boundary of $S$. The difference $S\setminus c$ is then a set of disconnected piecewise analytic surfaces $S_1,...,S_n$. Then taking the Weyl-operator $W_c:=W_S W_{S_1}^*...W_{S_n}^*$ corresponds to a flux through the 1-dimensional quasi-surface $c$. The analogue procedure can be used to construct Weyl-operators $W_x$ for 0-dimensional quasi-surfaces $x$. These 0-dimensional quasi-surfaces are viewed as fundamental in this paper and all other Weyl-operators are viewed as composites of these fundamental Weyl-operators (a similar approach is taken in \cite{sahlmann-thiemann}).

The gauge-invariant scaffold algebra arises as a restriction of the algebra of n-hand operators (as used e.g. in \cite{rovelli}), where all edges are required to be in the scaffold and the hands are vertices in the scaffold. Particularly, we consider observables of the form $Tr(W_{x_1}(\mu_1)h_{\gamma_{x_1}^{x_2}}W_{x_2}(\mu_2)...)$, where $\gamma_x^y$ connects the points $x$ and $y$ in the Cauchy surface. Due to the construction of the scaffold, there is at least one representative for each diffeomorphism class of graphs that consists only of elements of the scaffold.\footnote{As explained e.g. in \cite{rovelli}, one can construct the geometric operators from the n-handed operators alone.} Hence there is at least one scaffold representative for each diffeomorphism class of n-handed operators.

\subsection{Pairs of Embedding Maps}

The reason why we built the scaffold with ''straight'' edges in the previous subsection was to be able to explicitly calculate the dependence of the holonomy along these edges on the degrees of freedom of a homogeneous connection. We saw previously that the degrees of freedom of a homogeneous connection can be expressed as\footnote{This expression is again understood in the global chart, defined and used throughout the previous section.}:
$$
  A = \Lambda^I_a dx^a \tau_I.
$$
Given a straight edge $e=((i_1,i_2,i_3),(f_1,f_2,f_3))$, we can calculate the holonomy from $i^a$ to $f^a$ along this edge for a homogeneous connection:
\begin{equation}
  \begin{array}{rl}
  h_e(A)&=\mathcal P \biggl\{\exp\biggl(\int_e A \biggr)\biggr\}\\
        &=\exp\biggl(\int_0^1 dt (f^a-i^a) \Lambda^I_a dx^a \tau_I\biggr)\\
        &=\mathbb I \cos\biggl(\frac L 2\biggr)+ 2 \hat{n}^I\tau_I \sin\biggl(\frac L 2\biggr),
  \end{array}
\end{equation}
where we used the shorthand
$$
  \begin{array}{rl}
    L=&||(f^a-i^a)A_a||=\sqrt{\sum_I ((f^a-i^a)A_a^I)^2},\\
    \hat{n}^I=&\frac{(f^a-i^a)A^I_a}{L}=\frac{\hat{e}^aA^I_a}{L}.
  \end{array}
$$
Using this expression, we can calculate the explicit dependence of the holonomy along any edge in the scaffold on the homogeneous connection. All edges have fiducial length $n l_o$ and are piecewise straight. Thus, for each straight piece\footnote{The extra links $l_{abcn}$ are decomposed into three straight pieces for these considerations. We denote these straight $l^i_{abcn},l^m_{abcn}$ and $l^f_{abcn}$ for the initial, middle and final piece respectively, when the orientation is chosen in positive $x_1$-direction.}, there is a natural number $n_e$ and a unit vector $\hat{e}$, such that $L=n l_o \sqrt{\sum_I(\hat{e}^a A_a^I)^2}=n l_o ||eA^I||$, such that for the straight pieces in the scaffold correspond to:
$$
  \begin{array}{rl}
    L_{e^i_{abc}}=&         l_o ||A_i||\\
    L_{l^i_{abcn}}=&        l_o ||\hat{e}(\alpha(a,n))A^I||\\
    L_{l^m_{abcn}}=&        n l_o ||A_1||\\
    L_{l^f_{abcn}}=&        l_o ||\hat{e}(\alpha(a,n))A^I||.
  \end{array}
$$
We will introduce the shorthand $\tau(e,A)$ rather than wasting paper with explicitly calculating $\hat{n}$ for all edges $e$ in the scaffold and all homogeneous connections $A$ and observe that for the diagonal gauge\footnote{This requires a gauge-transformation and in general a diffeomorphism amounting to a rotation of our chart.}, where the connection takes the form $A=a dx^1\tau_1 + b dx^2 \tau_2 + c dx^3 \tau_3$ there is no dependence of $\tau(e)$ on $A$.

Let us now insert a connection $A_{LRS}(a,b,c)$ that is locally rotaionally symmetric around the $x_1$-axis as well as an isotropic connection $A_{iso}(c)$\footnote{The parameters $b,b$ and $c$ denote the degrees of freedom of the reduced connections as introduced in section \ref{red-conn}. We will choose a diagonal gauge for these connections, such that the connections simplify to $A_{LRS}=a(dx^1 \tau_1+dx^2 \tau_2)+c dx^3 \tau_3$ and $A_{iso}=cdx^i\tau_i$ respectively.}.

The matrix elements of the holonomies $h_{e^i_{abc}}(A_{LRS})$ are easily calculated to be linear combinations of the exponentials $e^{i l_o/2 (m a+ n c)}$\footnote{All numbers denoted by $n,m$ are integers unless stated otherwise.}, where for $i=1,2$ we obtain a dependence on $e^{i l_o/2 m a}$ and for $i=3$ an dependence on $e^{i l_o/2 nc}$. Since the holonomies along the middle pieces of the extra links are concatenations of holonomies along edges that are parallel to lattice edges in $x_3$-direction, we obtain that their matrix elements $h_{l^m_{abcn}}(A_{LRS})_{ij}$ are also only linear combinations of $e^{i l_o/2 nc}$. Similarly, we obtain that the matrix elements of the holonomies $h_{l^i_{abcn}}(A_{LRS})$ along the initial and final pieces of the extra links are linear combinations of $e^{i l_o/2 ma}$. Using the observation that the isotropic model arises, when the connection components $a=c$, we obtain that holonomies along all concatenations of these curves are of the form:
\begin{equation}
  \begin{array}{rl}
    h_{scaffold}(A_{LRS})=&\sum_{nm} \xi_{nm} e^{i l_o/2 (ma+nc)}\\
    h_{scaffold}(A_{iso})=&\sum_n \xi_n e^{i l_o/2 nc}.
  \end{array}
\end{equation}
We have not included $h_{l^f_{abcn}}(A_{LRS})$ and $h_{l^f_{abcn}}(A_{iso})$, because these are due to homogeneity and the groupoid morphism structure of the connection $h_{l^I_{abcn}}^{-1}(A_{LRS})$ and $h_{l^f_{abcn}}^{-1}(A_{iso})$ respectively.

We notice that all matrix elements of holonomies along scaffold edges for $A_{LRS}$ connections are linear combinations of $e^{i l_o/2(n a + m c)}$ for some $n,m\in \mathbb Z$. The isotropic connection arises as the special case, when $a=c$, i.e. the matrix elements of holonomies simplify to linear combinations $e^{i l_o/2 n c}$. This simplification will not occur if we consider a diagonal homogeneous connection $A_{diag}=a dx^1\tau_1+b dx^2 \tau_2+c dx^3\tau_3$, because the ''legs'' ($l^i_{abcn}$ and $l^f_{abcn}$) of the extra legs have can not be expanded in $e^{i l_o/2(n_1a+n_2b+n_2c)}$, because of the extra square root appearing in the $L$s.

The technical difficulties arising from this fact are the reason, why we postpone the general homogeneous case to later work.

Let us now construct the map $P$ for the LRS and isotropic model by defining it on an arbitrary cylindrical function $F_\gamma(h_{e_1},...,h_{e_n})$. As we outlined in our strategy, we will proceed in two steps: First we use diffeomorhpism invariance of the kinematical states in the full theory to map the graph $\gamma$ onto the scaffold. For practical reasons, we will do this by hand for some simple graphs, while we will refer to the construction outlined in the previous subsection for general graphs.

Let us start with the general case: Given a graph $\gamma$, we can use the construction of an embedding of this graph into the scaffold, which has ambiguities, because we did not specify the details of the embedding of the braids $B_1$ and $B_2$ into the regular lattice. However using the axiom of choice, there exists at least one embedding of the concatenated braids that fits into a minimal cube $C_n=\{\varphi(a,b,c)\in \Sigma:-n l_o \le a,b,c \le n l_o; n \in \mathbb N\}$ around the origin of our chart. Thus, for each graph $\gamma$ there exists a minimal number $\mathcal N(\gamma)$ such that, by using our construction, $\gamma$ is embedable into the cube $C_{\mathcal N(\gamma)}$. The embedding of $\gamma$ into the scaffold restricted to $C_{\mathcal N(\gamma)}$ will however in general not be unique, so there exists a set $\mathcal I(\gamma)$ of embeddings of gamma into the scaffold region $C_{\mathcal N(\gamma)}$.

We have not yet considered the orientation of the edges in $\gamma$. Given an edges $e$ out of a graph $\gamma$, our procedure may embed the edge by an extended diffeomorphism $\phi$ oriented or with reverse orientation into the scaffold. If the edge $e$ is reversed, then we will make use of the fact that a quantum connection is a groupoid morphism form the path groupoid into the gauge group and alter the cylindrical function $F(h_{e_1},...,h_{e_n})$ to:
\begin{equation}
  \tilde{F}(h_{e_1},...,h_{e_n}):=F(h^{sgn(\phi(e_1))}_{e_1},...,h^{sgn(\phi(e_n))}_{e_n}),
\end{equation}
where the sign of $\phi(e)$ is positive if $\phi$ maps $e$ into an edge with the same orientation and negative if $e$ is mapped into opposite orientation.

This lets us define a map $P_o$, that assigns to each cylindrical function $F_\gamma$ the average over the embeddings into the cube $C_{\mathcal N(\gamma)}$:
\begin{equation}
  P_o: F_\gamma \mapsto \frac{1}{|I(\gamma)|}\sum_{\phi\in\mathcal I(\gamma)} \tilde{F}_{\phi(\gamma)}.
\end{equation}
Notice that we just needed to employ covariance under extended diffeomorphisms for this construction, since for any cylindrical function this procedure amounts to average over the action of a set of diffeomorphisms on it. This does change the cylindrical function, however since we are interested in constructing a diffeomorphism invariant theory, this is an entirely allowed step, because it amounts to gauge fixing the diffeomorphisms.

Using $P_o$, we define the quantum reduction map $P$ by restricting the image of $P_o$ to symmetric connections. This means for a cylindrical function $F_\gamma$:
\begin{equation}
  P : F_\gamma \mapsto (P_o \tilde{F}_\gamma)\biggl|_{A_{sym}}\biggr.
\end{equation}
For technical reasons that become obvious in the construction of $i$ it turns out that it is more convenient to gauge fix the diffeomorphisms for some simple graphs ''by hand'' (using a map $P_s$ for a set $\mathcal S$ of ''simple'' knot classes of graphs), thus modifying the definition of $P$ to $P : F_\gamma \mapsto (P_s \tilde{F}_\gamma)\biggl|_{A_{sym}}\biggr. \forall \gamma \in \mathcal S$ and $P : F_\gamma \mapsto (P_o \tilde{F}_\gamma)\biggl|_{A_{sym}}\biggr.$ for $\gamma$ otherwise. We will denote the corresponding maps for $A_{sym}$ consisting of LRS-connections by $P_{LRS}$ and the case of an isotropic connection by $P_{iso}$.

Let us now calculate the image of an arbitrary Spin network function $T_\gamma$ under $P_{iso}$. $T_\gamma$ has the structure $T_\gamma(A)=\Pi_{e\in E(\gamma)} \rho^{i_e}(h_e(A))_{n_em_e}$, where $\rho^i(g)_{mn}$ denotes the matrix element $n,m$ of the $i$th representation of $g$. Let us use the unitarity of the matrix representations, i.e. $\rho^i(h^{-1}_e)_{nm}=\overline{\rho^i(h_e)_{mn}}$, which means that $\tilde{T}_\gamma$ is of the same form as $T_\gamma$ except for possible complex conjugations. However we have already observed that matrix-elements of the holonomies along edges in the scaffold reduce to linear combinations of $e^{i n l_o/2 c}$ for isotropic connections. Moreover, using the representation theory of $SU(2)$, which states that $\rho^i(g)_{mn}$ can be constructed by symmetrizing $i$ products of matrix elements of the fundamental representation of $g$, we obtain that each spin-network function $T_\gamma$ is reduced to a linear combination of $e^{i l_o/2 n c}$:
\begin{equation}
  P_{iso}(T)=\sum_{n\in D \subset \mathbb Z} \xi_n e^{i l_o/2 n c},
\end{equation}
where $D$ is a finite set. Since $P$ is a linear operation and the spin-network functions are dense in the cylindrical functions, we see that the cylindrical functions lie in the completion:
\begin{equation}
   P_{iso}(Cyl_o)=\sum_{n\in \mathbb Z} \xi_n e^{i l_o/2 n c}
\end{equation}
with possibly infinitely many summands and the usual restrictions on the $\xi_n$ for Fourier coefficients of continuous functions on $0,...,2\pi$. Obviously the same arguments holds for $P_{LRS}$ with only a little more notational effort. Thus, for any cylindrical function $Cyl$ we have:
\begin{equation}
  P_{LRS}(Cyl_o)=\sum_{n,m\in\mathbb Z} \xi_{n,m} e^{i l_o/2(ma+nc)}.
\end{equation}
Using the fact that the set $\{e^{i l_o/2(ma+nc)}:n,m\in\mathbb Z\}$ is dense in the image of $P_{LRS}$, we can use the Fourrier transform to extract the coefficients $\xi_{nm}$:
\begin{equation}
  \xi_{nm}(Cyl_o) = \int_0^\frac{ \pi }{l_o} \frac{dc}{2 \pi} \int_0^\frac {\pi}{ l_o} \frac{da}{2 \pi} P_{LRS}(Cyl_o) \exp(-il_o/2(am+cn)).
\end{equation}
With these coefficients it is convenient to write $P(Cyl_o)=\sum_{nm} \xi_{nm} e^{il_o/2(am+cn)}$.

With the observation that the $e^{i l_o/2(am+cn)}$ span the image of $P_{LRS}$, we can easily construct a map linear map $i$ from the image of $P_{LRS}$ into the cylindrical functions such that $P\circ i(e^{i l_o/2(am+cn)})=e^{i l_o/2(am+cn)}$. Let us start with the isotropic case. Using the standard $\tau^3=i/2 diag(1,-1)$, we can write the holonomy along $e^3_{0,0,0}$ as:
\begin{equation}
  h_{e^3_{0,0,0}}=\biggl(
    \begin{array}{cc}
      e^{i l_o/2 c} & 0 \\ 0 & e^{-i l_o/2 c}
    \end{array}
  \biggr),
\end{equation}
which tells us that $e^{i l_o/2cn}$ can be written as:
\begin{equation}
  e^{i l_o/2cn}=\left(\left(h_{e^3_{0,0,0}}(A)\right)_{11}\right)^n\biggl|_{A_{iso}(c)} \biggr.
\end{equation}
Let us now define the set $\mathcal S_{iso}$ of simple knot classes for isotropic graphs as the set of graphs that consist only of one edge. For these we define $P_{s,iso}$ to be the map:
\begin{equation}
  P_{S,iso} : F(h_e) \mapsto F(h_{e^3_{0,0,0}}).
\end{equation}
With these preparations, we have a pair $P_{iso},i_{iso}$ that constitutes a quantum embedding, if $i_{iso}$ is defined as:
\begin{equation}
  i_{iso} : e^{i l_o/2 cn} \mapsto \left(\left(h_{e^3_{0,0,0}}\right)_{11}\right)^n.
\end{equation}
The map $i_{iso}$ is the extension by linearity to the span of $e^{i l_o/2 nc}$, meaning that $$i_{iso}: \sum_n \xi_n e^{i l_o/2nc} \mapsto \sum_n \xi_n \left(\left(h_{e^3_{0,0,0}}\right)_{11}\right)^n.$$

The construction for the LRS-reduction is completely analogous: All that we have to change is that we have to extend our definitions to account for the $a$-dependence of $e^{i l_o/2(ma+nc)}$. Let us consider the holonomy along $e^1_{1,0,0}$, which depends on the LRS-connection as:
\begin{equation}
  h_{e^1_{1,0,0}}=\mathbb I \cos(\frac {l_o} 2 a) + 2 \tau^1 \sin(\frac {l_o} 2 a)=
  \biggl(  \begin{array}{cc}
    \cos(\frac {l_o} 2 a) & i \sin(\frac {l_o} 2 a) \\
    i \sin(\frac {l_o} 2 a) & \cos(\frac {l_o} 2 a)
  \end{array} \biggr),
\end{equation}
from which we deduce that
\begin{equation}
  e^{i l_o/2 m a} = \left((h_{e^1_{1,0,0}})_{11}+(h_{e^1_{1,0,0}})_{12}\right)^m \biggl|_{A_{LRS}(a,c)}\biggr..
\end{equation}
Extending the set $\mathcal S_{iso}$ by including not only graphs depending on only one edge but also graphs that depend on three unconnected edges we obtain the set $\mathcal S_{LRS}$. For these two knot classes of graphs, we extend the map $P_{S,iso}$ to graphs with two unconnected edges and define $P_{S,LRS}$ as:
\begin{equation}
  P_{S,LRS} : \biggl\{
  \begin{array}{ll}
    F(h_e)  & \mapsto F(h_{e^3_{0,0,0}})\\
    F(h_{e_1},h_{e_2},h_{e_3}) & \mapsto F(h_{e^3_{0,0,0}},h_{e^1_{1,0,0}},h_{e^3_{2,0,0}})
  \end{array}\biggr..
\end{equation}
This defines $P_{LRS}$, however the definition of $i_{LRS}$ needs the construction $e^{i l_o/2 a}$ through a cylindrical function, which can be done as follows:
\begin{equation}
  e^{i l_o/2 a} = \biggl((h_{e^3_{0,0,0}})_{11}(h_{e^3_{2,0,0}})_{22}\left((h_{e^1_{1,0,0}})_{11}+(h_{e^1_{1,0,0}})_{12}\right)\biggr)\biggl|_{A_{LRS}(a,c)}\biggr..
\end{equation}
This lets us define $i_{LRS}$ as the extension by linearity of:
\begin{equation}
  \begin{array}{l}
  i_{LRS}: e^{i l_o/2 (ma+nc)} \mapsto \\
 \biggl((h_{e^3_{0,0,0}})_{11}(h_{e^3_{2,0,0}})_{22}\left((h_{e^1_{1,0,0}})_{11}+(h_{e^1_{1,0,0}})_{12}\right)\biggr)^m \left(\left(h_{e^3_{0,0,0}}\right)_{11}\right)^n.
 \end{array}
\end{equation}
Although these two maps $i_{LRS}$ and $i_{iso}$ satisfy $P_{LRS}\circ i_{LRS}=Id_{Img(P_{LRS})}$ and $P_{iso}\circ i_{iso}=Id_{Img(P_{iso})}$, it will later turn out to be useful to formulate these maps in terms of spin-network functions. Using $(h_{((-2m,0,0),(-m,0,0)})_{11}(A_{LRS}(a,c)=e^{i l_o/2 am}$ and $(h_{((0,0,-2n),(0,0,-n)})_{11}(A_{LRS}(a,c)=e^{i l_o/2 cn}$ we alter the maps to:
\begin{equation}
  \begin{array}{rcl}
    i_{LRS}: & e^{i l_o/2 (ma+nc)} \mapsto & (h_{((-2m,m,0),(-m,m,0)})_{11} (h_{((n,n,-2n),(n,n,-n)})_{11} :=T_{n,m}\\
    i_{iso}: & e^{i l_o/2 nc} \mapsto & (h_{((n,n,-2n),(n,n,-n))})_{11}:=T_n
  \end{array}
\end{equation}
These edges where chosen in such a way that they do not intersect each other for different $n,m$.

These two quantum embedding maps $(P_{iso},i_{iso})$ and $(P_{LRS},i_{LRS})$ will be used in the next section to extract cosmological sectors form Loop Quantum Gravity. We will call them {\bf microscopic embedding}, because they are constructed to preserve the microscopic structure of the graphs.

\subsection{Comments about the Embedding Maps}

Several remarks are in order:\\
$\bullet$ The construction of the scaffold defines an embedding of the combinatorial groupoid into the path groupoid of $\sigma=\mathbb R^3$ in the obvious way. We saw on the other hand that the restriction to the scaffold allows us to define cylindrical functions that are diffeomorphism fixed, when we apply an averaging over the allowed embeddings into the scaffold. This displays the relation that the combinatorial theory and the diffeomorphism fixed theory play: The averaging over the allowed embeddings into the scaffold simply amounts to gauge fixing the action of a subgroup of the automorphism group of the scaffold, which is precisely the subgroup that is generated by the action of homeomorphisms of the base manifold $\sigma$.

$\bullet$ At the beginning of this work, we considered not only the microscopic embedding that we showed here, but also embeddings in which the image of $i$ takes a certain form. We thought e.g. that an embedding in which $i$ maps into complexifier coherent states (see \cite{gcs}) would result in a different ''semiclassical embedding''. However, it turned out that the reduced algebra  as well as the induced representation was not different from the one constructed with the microscopic embedding map as long as the cylindrical functions in the image of $i$ had been built on the same graphs. The significant dependence of the reduced system on the details of the graph will have significant (e.g. different dynamics) consequences as we will discuss in section \ref{dynamics}.

$\bullet$ We have already argued for geometrical homogeneity. The construction of such states is however, again due to the graph dependence of this statement, very difficult. This difficulty was the main reason for proceeding in analogy to \cite{abl}. After the completion of this piece of research and before its publication we discovered that there are indeed states on the algebra that underlies LQG, that satisfy an exact geometric homogeneity \cite{dqg}. These states do however not arise through a quantum symmetry reduction.

$\bullet$ One can argue that we would have obtained a completely different pair of embedding maps, if we used a different scaffold, particularly one that is not based on a regular lattice, but e.g. one that is constructed on a lattice that has fundamental length $l_o$ in $x_1$ and $x_2$ direction and an irrational multiple $l_1$ in $x_3$ direction. The corresponding locally rotational symmetric embedding would not differ much, because the image of $P$ would contain linear combinations of $e^{i/2 (l_o m a + l_1 n c)}$, which is isomorphic to the image of $P_{LRS}$ in our construction. The isotropic model would however differ significantly, because the image of $P$ would contain $e^{i/2 (l_o m c + l_1 n c)}$, which is different form the image of $P_{iso}$. The resulting isotropic model would be similar to \cite{velhinho}, which has the Bohr compactification of $\mathbb R$ as its quantum configuration space.

A even more significantly different embedding can be constructed in the following way: consider a scaffold that is constructed as in this section, but based on a lattice that has an infinite number of mutually irrational length $l_i$. Such a lattice can be constructed from the regular lattice by applying the diffeomorphism that has the following form in the lattice chart:
\begin{equation}
  \phi: (x_1,x_2,x_3) \mapsto (x_1,x_2,\sinh(x_3)).
\end{equation}
The the image of the map $P$ that is built on a scaffold based on such a lattice will contain linear combinations of $e^{i/2 l_o (m c + \sum_{k} n_k \sinh(k) c)}$ and thus a basis will be labeled by an infinite number of integers $m,n_k$. The induced Hilbert space representation of the reduced theory is however in general not unitarily equivalent to he standard representation of LQC on $L^2(\bar{\mathbb R}_{Bohr})$. We will not discuss the physical implications of these more complicated quantum embeddings, and rather focus on the simplest case in the next section.

\section{Embedable Loop Quantum Cosmology}

Having the reduction maps at our disposal, we will apply them to extract a cosmological sector from LQG.

\subsection{Preparatory Considerations}

The method of embedding an algebra using a quantum embedding $(P,i)$, as we presented it in \cite{reduction}, requires the existence of a Hilbert-$C^*$-module for the algebra of the full system. However, such a construction is unfeasible for the holonomy-flux-Weyl-algebra, because any Weyl-operator that corresponds to a flux through a finite 2-dimensional quasi-surface will affect an over-countable number holonomies (holonomies along all edges that intersect transversely with the surface) and thus affect an over-countable number of cylindrical functions, which can not be taken care of by relations between countably many cylindrical functions, as they arise from Hilbert-$C^*$-modules. This shortcoming can be cured by considering a theory in which the diffeomorphisms are gauge fixed. We have already considered a gauge fixing of the diffeomorphisms for cylindrical functions. A general bounded elementary operator of Loop Quantum Gravity will however be constructed from cylindrical functions and Weyl-operators, i.e. it is of the form:
\begin{equation}
  O=\sum_i Cyl^i_{\gamma_i} W_{S^i_1}(\lambda^i_1)...W_{S^i_n}(\lambda^i_n),
\end{equation}
where $W_{S,f}(\lambda)$ denotes the Weyl-operator $\exp(i \lambda E(S,f))$ associated to the flux $E_i(S)$ through the surface $S$ and where $Cyl$ is a cylindrical function. Since any operator product of these elementary operators $O$ will again be of this form, we conclude in analogy to the gauge fixing of the diffeomorphisms on cylindrical functions that we have to gauge fix the diffeomorphisms by assigning exactly one representative $(\gamma,S_1,...,S_n)$ for each class of topological relations that the edges $e\in\gamma$ can have with each other (i.e. the knot class of the graph) and with the surfaces $S_1,...,S_n$ (i.e. the transversal intersections of the edges in $\gamma$ have to be left invariant under the gauge fixing of the diffeomorphisms). Due to the over-countability issue, we will at the end view Weyl-operators $W_x$ on 0-dimensional quasi-surfaces $x$ as elementary operators. This seems a very weak restriction as we explained before. In order to explain the additional difficulties that arise from the nontrivial topological relations arising from 1- and 2-dimensional quasi-surfaces, we will consider a construction that fixes their diffeomorphism class:

Given a set $(\gamma,S_1,...,S_n)$, one can find such a construction as follows\footnote{We will assume that all $S_i$ are homeomorphic to a disc. These are sufficiently many surfaces to construct all operators that are necessary to discuss Loop Quantum Gravity.}:
\begin{enumerate}
  \item For each intersection of a surface with an edge in $\gamma$ we split this edge and add a bi-valent vertex. The resulting graph is denoted $\tilde{\gamma}$. Intersections of surfaces with $\gamma$ at vertices are left untouched.

  \item We use the construction described in the previous section for $\tilde{\gamma}$ to embed it into the scaffold. We have to fix the choice of embedding the braids $B_1$ and $B_2$ for each individual graph $\tilde{\gamma}$ to make the embedding unique.

  \item Since $\tilde{\gamma}$ intersects with the surfaces only at vertices, there is a piecewise analytical ''small surface'' that has the same topological relations with the edges of the graph as a small neighborhood of the intersection of the surface with graph does. This means in particular that it intersects the graph only at this vertex. We denote these ''small surfaces'' by $S_{i,v}$, where $i$ denotes the index of the surface and $v$ the vertex in $\tilde{\gamma}$.

  \item The surfaces $S_{i,v}$ can now be connected using surfaces that do not intersect with $\tilde \gamma$, because the scaffold has no accumulation point of edges.\footnote{If we label the vertices that intersect $S_i$ by an index $j$, then there are ''small bands'' $B_j$ that connect the vertex $j$ with $j+1$. Then the concatenation of $S_{iv_1},B_1,S_{iv_2},...,S_{iv_N}$ has the desired topological relations and is homeomorphic to a disc.} Moreover, we can choose these connecting pieces to be piecewise analytic.

\end{enumerate}
The embedding of $\gamma,S_1,..,S_n$ is still dependent on the labeling of the edges, vertices and graphs. To get rid of this dependence in the gauge fixing of our operators $O$, we have to average over all possible embeddings of $\gamma,S_1,...,S_n$ into the diffeomorphism fixed set $\tilde \gamma,\tilde S_1,...,\tilde S_n$:
\begin{equation}
  \tilde{O} = \frac 1{|\mathcal I|} \sum_{i \in \mathcal I} i^*(O),
\end{equation}
where $\mathcal I$ denotes the set of all possible embeddings and $i^*(O)$ denotes the pull-back of the expression for $O$ under these embeddings. We conclude that for each elementary operator $O$ depending on $(\gamma,S_1,...,S_n)$, we can associate a diffeomorphism fixed operator $\tilde{O}$ using the above construction. The relation of the diffeomorphism fixed theory to Loop Quantum Gravity is as follows:
\begin{enumerate}
  \item Loop Quantum Gravity can be though of as the quantum field theory that is generated by the bounded elementary operators of the form $O=\sum_i Cyl^i_{\gamma_i} W_{S^i_1}(\lambda^i_1)...W_{S^i_n}(\lambda^i_n)$ and its Hilbert-space representation arises as the GNS-representation using the Schr\"odinger functional:
  \begin{equation} \label{schr-fun}
    \omega_S(O)=\int d\mu_{AL}(A) \sum_i Cyl^i_{\gamma_i}(A).
  \end{equation}
      Notice that this functional is invariant under the choice of gauge fixing of the diffeomorphisms.
  \item Consider the diffeomorphism gauge fixed theory that can be define as follows: Given a set of diffeomorphism fixed operators $\tilde O_1,...,\tilde O_m$, there is a graph $\gamma$ (the union of all occurring edges and vertices in the individual graphs) and a smallest set $\{S_1,...,S_k\}$ of surfaces (also the union of all occurring surfaces). This means we can define the operator product $\tilde O_1...\tilde O_m$ on this set\footnote{To check that this operator product is well defined one needs to check its associativity, which is obvious from the construction.}. Now use the fact that the Schr\"odinger functional is invariant under the extended diffeomorphisms, which allows us to use it to preform a GNS construction on this combinatorial theory.
  \item To simplify our discussion, we make the observation that we can take any vertex $v$ in this graph and any edge $e$ originating in this vertex and find a small piecewise analytic surface $S_{v,e}$ such that $e$ is above the surface\footnote{We have not mentioned the orientation of surfaces yet, because it was not necessary in the construction of the Weyl-operators. This is due to the fact that an inversion of the orientation can be absorbed in a flip of the sign of the $\lambda$-parameter in the Weyl-operator $W_S(\lambda)$.} and all other edges are parallel to this surface and exit in parallel to it. We notice that products of the Weyl-operators $W_{S_{v,e}}(\lambda)$, where $e$ and $v$ range over the entire graph, are able to construct the action of any Weyl-operator. Thus, we can write the bounded elementary operators as $O=\sum_i Cyl^i_{\gamma_i} W_{S^i_{v1,e1}}(\lambda^i_1)...W_{S^i_{v_n,en}}(\lambda^i_n)$\footnote{There is a subtlety which is the precise mathematical reason for restricting our attention to the Weyl-operators on 0-dimensional surfaces: The operators $\sum_i Cyl^i_{\gamma_i} W_{S^i_{1,v1,e1}}(\lambda^i_1)...W_{S^i_{n,vm,em}}(\lambda^i_n)$ are not symmetric under difeomorphisms mapping the set $\gamma,S_1,...,S_n$ onto itself. If we want to achieve this by averaging over all topological relations, then we need to recover the original surfaces $S_i$. This means that we would have to introduce additional structure on the Hilbert-$C^*$-module that would encode these topological relations. It is however not obvious how to model this structure mathematically on the Hilbert-$C^*$-module.}. This means that we can consider the combinatorial theory as a covariant pair consisting of the commutative algebra of cylindrical functions and the group of elementary Weyl-operators $W_{S_{v,e}}(\lambda)$. These elementary operator can not be distinguished from the ''elementary'' Weyl-operators $W_x$ on 0-dimensional quasi-surfaces $x$  and the additional bands to connect these elementary Weyl-operators have no effect on the scaffold holonomies and can thus be replaced with unit operators.
  \item Notice that the one can put the previous point (2.) on its head and construct any bounded elementary operator $O$ form a gauge-fixed operator $\tilde O$ and an extended diffeomorphism. We conclude that Loop Quantum Gravity can be constructed from the combinatorial theory and the extended diffeomorphisms, since the Schr\"odinger functional of the combinatorial theory is invariant under these extended diffeomorphisms. Moreover Loop Quantum Gravity whose elementary Weyl-operators are on 0-dimensional quasi-surfaces can be constructed from an embedding of the scaffold algebra.
\end{enumerate}
The combinatorial theory is no longer plagued by the over-countability issue of the action of the Weyl-operators. Thus, we are able to construct a Hilbert-$C^*$-module over this combinatorial theory and then use the quantum embeddings that we defined in the previous section to extract cosmological sectors from this combinatorial theory. This sector is then by construction a sector of the diffeomorphism invariant Loop Quantum Gravity, due to the connection that we just described.

\subsection{Application of the microscopic Embedding}

The first step in the construction of a quantum embedding is the construction of a Hilbert-$C^*$-module for the combinatorial theory that we constructed in the previous section. It is particularly simple to construct a Schr\"odinger-type Hilbert-$C^*$-module for a covariant pair that consists of the commutative algebra of continuous functions on a compact space\footnote{The configuration space of the combinatorial theory is compact because it is a closed subspace of the configuration space of Loop Quantum Gravity, which is a compact space.} and a group of homeomorphisms acting freely and properly thereon. Due to the previously explained relation of the scaffold algebra and the diffeomorphism fixed algebra of LQG with Weyl-operators on 0-dimensional quasi-surfaces, we will use a Hilbert-$C^*$-module for the scaffold algebra. The cylindrical functions on the scaffold furnish a pre-Hilbert-pre-$C^*$-module by construction.

It actually suffices to construct a pre-Hilbert-$C^*$-module, because the completion of these is unique. Using the fact that the diffeomorphism fixed cylindrical functions are dense in the algebra of continuous functions on the diffeomorphism fixed configuration space allows to use diffeomorphism fixed cylindrical functions to construct a Hilbert-$C^*$-module for the combinatorial theory. Let us consider the module set that consists of diffeomorphism fixed cylindrical functions: $F_\gamma=\frac 1 {|\mathcal I|}\sum_{\phi \in \mathcal I}Cyl_{\phi(\gamma)}$. We use the bilinear structure:
\begin{equation}
  \begin{array}{rl}
  \langle F^1_\gamma,F^2_\gamma \rangle^\circ_A :=& F^1_\gamma \int \Pi_{e \in E(\gamma)} d\mu_H(g_e) \overline{F^2(g^{-1}_{e_1}h_{e_1},...,g^{-1}_{e_n}h_{e_n})} \\ & W_{S_{e_1}}(\lambda(g_{e_1}))...W_{S_{e_n}}(\lambda(g_{e_n})).
  \end{array}
\end{equation}
$S_{e_i}$ is a shorthand for the 0-dimensional quasi-surface $S_{s(e_i)e_i}$, that has only $e_i$ as an above edge and all other edges at $s(e_i)$ are inside, so the Weyl-operator acts trivially on them. $d\mu_H(g)$ denotes the Haar measure on $SU(2)$ over a copy whose representatives are denoted by $g$. $\lambda(g)$ denotes a function on the group $SU(2)$ with values in the Lie-algebra, such that $\exp(\lambda (g)\tau)=g$.

This bilinear structure is not yet symmetric under the automorphisms of $\gamma,S_1,...,S_n$, so we have to symmetrize over all embeddings $i\in \mathcal I$ of $\gamma,S_1,...,S_n$ into $\tilde\gamma,\tilde{S}_1,...,\tilde S_n$ that preserve the topological relations\footnote{Notice that the $S_i$ denote vertices in the scaffold that point at precisely one adjacent edge and not generic quasi-surfaces.}. This yields the final form of the bilinear structure $\langle .,. \rangle_A$ as:
\begin{equation}
  \langle F_1, F_2 \rangle_A := \frac{1}{|\mathcal I|} \sum_{i \in \mathcal I} \langle F_1, F_2 \rangle_{A}.
\end{equation}
A proof that this bilinear structure is indeed dense in the combinatorial theory of diffeomorphism fixed LQG is given by applying the analogue of the approximate identity that we used to construct the scaffold algebra and preforming the symmetrization over the equivalent topological relations.

Using this Schr\"odinger-type pre-Hilbert-$C^*$-module, we can use the quantum embedding $(P,i)$ for cylindrical functions $F_\gamma$ immediately, because:
\begin{equation}
  \begin{array}{rcl}
    P: & F & \mapsto P(F)\\
    i: & h & \mapsto i(h),
  \end{array}
\end{equation}
are well defined. Applying the rules for quantum embeddings, we construct the reduced module as the space $P(F_\gamma)$, where $F_\gamma$ ranges over all cylindrical functions. Using the action of the operators $\langle F_1,F_2 \rangle_A H:= \langle F_1,F_2 \rangle_A H$, we define the bilinear structure for all $P(F)$ as:
\begin{equation}
  \langle P(F_1), P(F_2) \rangle_B P(H):= P(\langle i(P(F_1)), i(P(F_2)) \rangle_A i(P(H))).
\end{equation}
Let us now insert the specific quantum embeddings $(P_{LRS},i_{LRS})$ and $(P_{iso},i_{iso})$ to obtain the specific quantum embeddings:

We learned in the previous section that the image of $P_{LRS}$ is spanned by the functions $e^{i l_o/2 (am+cn)}$, where $(a,c)$ denote the degrees of freedom of the LRS-model in diagonal gauge. Thus, for any pair $(F,f)$, we obtain the corresponding element of the embedded module:
\begin{equation}
  P_{LRS}(F)=\sum_{n,m} \xi_{n,m} \exp(i l_o/2 (am+cn)).
\end{equation}
We will calculate the bilinear structure by the action of pairs $e^{i l_o/2 (am+cn)}$ on the dense set $e^{i l_o/2 (am+cn)}$ and consider the completion by sesquilinearity in the arguments and linearity in the representation space:
\begin{equation}
  \begin{array}{l}
  \langle e^{i l_o/2 (am_1+cn_1)},e^{i l_o/2 (am_2+cn_2)}\rangle_{LRS} e^{i l_o/2 (am_3+cn_3)}\\
  =P(\langle i(e^{i l_o/2 (am_1+cn_1)}),i(e^{i l_o/2 (am_2+cn_2)})\rangle_{A} i(e^{i l_o/2 (am_3+cn_3)}))\\
  =\delta_{m_2,m_3} \delta_{n_2,n_3} e^{i l_o/2 (m_1a+n_1c)},
  \end{array}
\end{equation}
where we used the observation that
$$
\begin{array}{l}
\int  d\mu_H(g_{((-2m,m,0),(-m,m,0))}) d\mu_H(g_{((n,n,-2n),(n,n,-n))})  \\
  T^*_{n_2m_2}( g^{-1}_{((-2m,m,0),(-m,m,0))}h_{((-2m,m,0),(-m,m,0)}, \\
  g^{-1}_{((n,n,-2n),(n,n,-n))}h_{((n,n,- 2n),(n,n,-n))}) \\
W_{S_{((-2m,m,0),(-m,m,0))}}(\lambda(g_{((-2m,m,0),(-m,m,0))}))\\
W_{S_{((n,n,-2n),(n,n,-n))}}(\lambda(g_{((n,n,-2n),(n,n,-n))}))
T_{n_3,m_3}
\end{array}
$$
reduces to $\delta_{n_2,n_3}\delta_{m_2,m_3}$ due to the action of the Weyl-operators as translations on $T_{n_3,m_3}$ if and only if $n_2=n_3$ and $m_2=m_3$, the translation invariance of the Haar measure and orthogonality of the spin-network-functions in the translated inner product w.r.t the Haar measure.

These relations $\langle (n_1,m_1),(n_2,m_2)\rangle_A (n_3,m_3)=\delta_{n_2,n_3}\delta_{m_2,m_3} (n_1,m_1)$ can immediately be identified with the quantum algebra of a particle on a torus, which we called $U(1)^2$ in \cite{reduction}. We can indeed identify the completion of $Img_{P_{LRS}}$ as functions on the torus and the bilinear structure as the rank-one-operators on the torus.

Let us now calculate the induced representation for this reduced quantum algebra: We have the association of ''rank one operators'' of the full and reduced theory:
\begin{equation}
  \langle F_1,F_2 \rangle_A \leftrightarrow \langle P(F_1),P(F_2) \rangle_{LRS}. \label{assoc}
\end{equation}
The Schr\"odinger ground state of Loop Quantum gravity yields for the ''rank one operators'' of the full theory:
\begin{equation}
  \omega(\langle F_1,F_2 \rangle_A) = \int d\mu_{AL}(A) \overline{F_2}(A) F_1(A).
\end{equation}
This implies for the the reduced rank-one operators:
\begin{equation}
  \begin{array}{l}
  \omega_{ind} (\langle e^{i l_o/2 (m_1a+n_1c)},e^{i l_o/2 (m_2a+n_2c)}\rangle_{LRS})\\
  =\omega (\langle i(e^{i l_o/2 (m_1a+n_1c)}),i(e^{i l_o/2 (m_2a+n_2c)})\rangle_A)\\
  =\delta_{m_1,o}\delta{m_2,o}\delta_{n_1,o}\delta_{n_2,o},
  \end{array}
\end{equation}
due to the orthogonality of the spin network functions $i(e^{i l_o/2 (ma+nc)})$ for different $m,n$ in the Schr\"odinger ground state of the full theory.

Let us compare this with the usual Schr\"odinger ground state on the torus:
\begin{equation}
  \omega_{torus}(f(a,c)W(\lambda_1\lambda_2))=\frac 1 {4\pi^2}\int_0^{2\pi}da\int_0^{2\pi}dc f(a,c),
\end{equation}
which yields for the rank one operators $\langle.,.\rangle_{torus}$ on the torus:
\begin{equation}
  \begin{array}{l}
  \omega_{torus} (\langle e^{i l_o/2 (m_1a+n_1c)},e^{i l_o/2 (m_2a+n_2c)}\rangle_{torus})\\
  =\frac 1 {4 \pi^2}\int_0^{2\pi}da\int_0^{2\pi}dc e^{i l_o/2 (a(m_1-m_2)+c(n_1-n_2))}
  =\delta_{m_1,o}\delta_{m_2,o}\delta_{n_1,o}\delta_{n_2,o},
  \end{array}
\end{equation}
which coincides with the induced ground state $\omega_{ind}$. Since the $\langle e^{i l_o/2 (m_1a+n_1c)},e^{i l_o/2 (m_2a+n_2c)}\rangle_{LRS}$ are a dense set of the operator algebra, we have demonstrated that (1) the LRS quantum algebra is the quantum algebra of a particle on the torus and (2) the induced representation is the canonical representation of this quantum system, which is unitarily equivalent to the Schr\"odinger representation on $L^2(\mathbb T,\frac 1{4\pi^2}da\wedge dc)$

Using the observation that the bilinear structure $\langle.,.\rangle_{iso}$ arises as the spacial case of the LRS-structure, where all $m_i=0$, which is due to the special analogy of our construction of $(P_{iso},i_{iso})$ compared to $(P_{LRS},i_{LRS})$, we can quote the result for the isotropic embedding without further calculations: $Img_{P_{iso}}$ is the module of functions on the circle and the bilinear structure yields the rank-one-operators on quantum mechanics on the circle. The induced representation is very analogous to the canonical representation of quantum mechanics on $U(1)$, which is unitarily equivalent to the Schr\"odinger representation on $L^2(U(1),d\mu_H)$. The gravitational part of the kinematics of standard LQC is equivalent to the canonical representation of the CCR-Weyl-algebra on $L^2(\bar{\mathbb R}_{Bohr})$ \cite{abl}, which contains infinitely many super selection sectors that can be reduced to the canonical representation of quantum mechanics on a circle on $L^2(U(1))$.

\subsection{Imposing Constraints}

Gravity is a constrained theory and Loop Quantum Gravity is a theory with three sets of constraints, the Gauss-constraint that generates ''ordinary gauge transformations'', the diffeomorphism constraint and the scalar constraint that generates something that are related to the ''timelike diffeomorphisms''. Since these transformations are not observable, one has to construct the quantum theory in a way that leaves these unobservable. This implies that any reduced theory of a constrained system has to be a reduced system, that is constructed as a reduction of the constraint surface. We will discuss the treatment of the Gauss- and the diffeomorphism constraint in this section and discuss the treatment of the scalar constraint under subsection \ref{dynamics}.

The solutions to the Gauss-constraint are products of traces of holonomies of closed loops. Thus,  we calculate the holonomies around closed loops in the scaffold. For an LRS-connection, these can all be generated by the three elementary loops\footnote{The loops involving extra links can be expressed as rotations of holonomies in the regular lattice due to local rotational symmetry of the connection.} $((0,0,0),(1,0,0),(1,1,0),(0,1,0),(0,0,0))$, $((0,0,0),(0,1,0),(0,1,1),(0,0,1),(0,0,0))$ and $((0,0,0),(0,0,1),(1,0,1),(1,0,0),(0,0,0))$:

$$
  h_{((000)(100)(110)(010)(000))}=\biggl(
  \begin{array}{cc}
    \cos (a\,l) +  \frac{1-i}{2} \,{\sin (a\,l)}^2 &
    2\left( i\,-1  \right) \,\cos (\frac{a\,l}{2})\,{\sin (\frac{a\,l}{2})}^3 \\
    2\left( 1+\,i  \right) \,\cos (\frac{a\,l}{2})\,{\sin (\frac{a\,l}{2})}^3 &
    \cos (a\,l) + \ \frac{1+i}{2}  \,{\sin (a\,l)}^2
  \end{array}
  \biggr)
$$
$$
  h_{((000)(010)(011)(001)(000))}= \biggl(
  \begin{array}{cc}
  {\cos (\frac{a\,l}{2})}^2 + \frac{{\sin (\frac{a\,l}{2})}^2}{e^{i \,c\,l}}&
   \frac{-\left( \left( -1 + e^{i \,c\,l} \right) \,\sin (a\,l) \right) }{2} \\
   \frac{\left( -1 + e^{-i \,c\,l} \right) \,\sin (a\,l)}{2}&
   {\cos (\frac{a\,l}{2})}^2 + e^{i \,c\,l}\,{\sin (\frac{a\,l}{2})}^2
   \end{array}
   \biggr)
$$
$$
  h_{((000)(001)(101)(100)(000))}= \biggl(
  \begin{array}{cc}
    \frac{1 - e^{i \,c\,l}\,\left( -1 + \cos (a\,l) \right)  + \cos (a\,l)}{2}&
   \frac{i }{2}\,\left( -1 + e^{i \,c\,l} \right) \,\sin (a\,l) \\
   \frac{\sin (a\,l)\,\sin (\frac{c\,l}{2})}{e^{\frac{i }{2}\,c\,l}} &
   \frac{1 - \frac{-1 + \cos (a\,l)}{e^{i \,c\,l}} + \cos (a\,l)}{2}
 \end{array}
 \biggr)
$$
The traces of these holonomies are all even periodic functions in both connection components $a,c$ with periodicity $l_o$:
\begin{equation}
  \begin{array}{rcl}
    Tr(h_{((000)(100)(110)(010)(000))})&=& 2 \cos(a l_o)+\sin^2(a l_o)\\
    Tr(h_{((000)(010)(011)(001)(000))})&=& 2 \cos^2(a l_o/2) +2 \cos(c l_o) \sin^2(a l_o/2)\\
    Tr(h_{((000)(001)(101)(100)(000))})&=& 2 \cos^2(a l_o/2) +2 \cos(c l_o) \sin^2(a l_o/2).
  \end{array}
\end{equation}
But the ''Wilson loops around these elementary plaquettes'' contain all the gauge invariant information of the homogeneous connection, because there are no ''smaller plaquettes'' in our scaffold. This means that all solutions to the Gauss constraint are even functions of periodicity $l_o$. It follows that we can solve the Gauss constraint for the reduced theory by restricting the domain of $P$ to gauge-invariant diffeomorphism-fixed spin network functions and by constructing $i$ in such a way that it takes values in gauge-invariant diffeomorphism-fixed functions only.

We have already calculated the image of the restriction of $P_{LRS}$ as the even functions of periodicity $l_o$. Now, we have to construct $i$ for these functions. This can be done by assigning to each $(2 \cos(a l_o)+\sin^2(a l_o))^n(2 \cos^2(a l_o/2) +2 \cos(c l_o) \sin^2(a l_o/2))^m$ the gauge-invariant and diffeomorphism fixed spin network-function:
\begin{equation}
  \begin{array}{rl}
    i_{LRS}: &(2 \cos(a l_o)+\sin^2(a l_o))^n(2 \cos^2(a l_o/2) +2 \cos(c l_o) \sin^2(a l_o/2))^m \\
    \mapsto &  (Tr(h_{((n00)(n+100)(n+110)(n10)(n00))}))^n(Tr(h_{((-mm0)(-mm+10)(-mm+11)(mm1)(-mm0))}))^m.
  \end{array}
\end{equation}
The linear extension to the entire image of $P_{LRS}|_{gauge.inv.}$ defines the gauge-invariant $i_{LRS}$.

Having a gauge-invariant pair $(P_{LRS},i_{LRS})$ enables us to calculate the reduced quantum algebra: We identify the action of the rank-one operators on the canonical module $(n,m)\sim(2 \cos(a l_o)+\sin^2(a l_o))^n(2 \cos^2(a l_o/2) +2 \cos(c l_o) \sin^2(a l_o/2))^m$, i.e. we consider
\begin{equation}
  \langle (n_1,m_1), (n_2,m_2) \rangle_{LRS} (n,m)=\langle i(n_1,m_1),(n_2,m_2) \rangle_A (n,m)=(n_1,m_1) \delta_{n_2,n}\delta_{m_2,m},
\end{equation}
where we reuse the observation that the action of the Weyl-operators reduces to a shifted integration of $\overline{i(n_2,m_2)}i(n,m)$ w.r.t. the Ashtekar-Lewandowski measure. The orthogonality of the spin-network-functions $i(n,m)$ for different $(n,m)$ then gives the result. The induced vacuum state $\omega_{ind}$ is the calculated by applying (\ref{assoc}) to the gauge invariant construction, which yields:
\begin{equation}
  \omega_{ind}(\langle (n_1,m_1),(n_2,m_2)\rangle_{LRS}):=\omega(\langle i(n_1,m_1),i(n_2,m_2)\rangle_A)=\delta_{n_1,o}\delta_{n_2,o}\delta_{m_1,o}\delta_{m_2,o}.
\end{equation}
This lets us define the intertwiner between the module spanned by $(n,m)$ and the Hilbert-module spanned by $e^{i(ax+by)}$ in $L^2(\mathbb T^2,\frac 1{4 \pi^2}dx\wedge dy)$, given by:
\begin{equation}
  U: (n,m) \mapsto e^{i(nx+my)},
\end{equation}
which is evidently unitary. Thus, we again obtain unitary equivalence of the reduced theory with quantum mechanics on the circle. To construct a unitary intertwiner is however more complicated than in the gauge-variant case.

Reusing the observation that one can obtain the isotropic model as the special case of the LRS model, where restrict ourselves to the states $(n,m)$ for which e.g. $m=0$ and to the operators that mediate between these, we obtain that the isotropic theory is unitarily equivalent to quantum mechanics on a circle. This is however a very weak result, because any two separable infinite dimensional Hilbert-spaces are unitarily equivalent. The physical interpretation of the underlying quantum configuration space is given by the intertwiner. This comes about as follows: The Hilbert-$C^*$-modules that we used to construct the quantum embedding where Schr\"odinger modules, i.e. the elements of the module have a pointwise multiplication defined among them. The lesson from noncommutative geometry is that the spectrum of the $C^*$-completion of this algebra is the desired topological space. However so far we have not been able to calculate this spectrum explicitly and are thus unable to present the gauge-invariant quantum configuratiion space of the LRS-model\footnote{We will continue to ask mathematicians until we find a solution.}.

{\it Considerations about the diffeomorphism constraint:} One can introduce the diffeomorphism invariance of Loop Quantum Gravity through the graph groupoid $\mathcal G$. The unit set of this groupoid consists of all embedded graphs $\gamma$ on the base manifold (in the Bianchi I setting $\mathbb R^3$). The morphisms of the groupoid consist of all ordered pairs $(\gamma_1,\gamma_2)$ of graphs that are in the same iso-knot class. The source map $s$ and the range map $r$ are given by:
\begin{equation}
  s(\gamma_1,\gamma_2)=\gamma_1 \,\,\, , \,\,\, r(\gamma_1,\gamma_2)=\gamma_2
\end{equation}
the inversion is $(\gamma_1,\gamma_2)^{-1}=(\gamma_2,\gamma_1)$, the composition law is:
\begin{equation}
  (\gamma_1,\gamma)\circ(\gamma,\gamma_2)=(\gamma_1,\gamma_2),
\end{equation}
and the object inclusion map is $e(\gamma)=(\gamma,\gamma)$. We can define an action $\alpha$ of this groupoid on the cylindrical functions $Cyl_\gamma$ using the momentum map $\mu$:
\begin{equation}
  \begin{array}{rl}
    \mu(Cyl_\gamma)=&\gamma\\
    \alpha((\gamma_1,\gamma_2),Cyl_{\gamma_2})=& Cyl_{\gamma_1}.
  \end{array}
\end{equation}
Calculating the orbits of this action of the graph groupoid on the cylindrical functions yields the equivalence classes $[Cyl_\gamma]_\sim$ of cylindrical functions taken w.r.t the equivalence relation: $Cyl^1_{\gamma_1}\sim Cyl^2_{\gamma_2}$ iff $Cyl^1=Cyl^2$ and $\gamma_1$ is an isomorphic to $\gamma_2$ as a knot.

The embedding procedure $P_o$ that we used to embed any given cylindrical function into the cylindrical functions defines equiavlence classes $[.]_P$ through:
\begin{equation}
  Cyl^1_{\gamma_1} \sim_P Cyl^2_{\gamma_2} \,\,{\textrm{ iff: }} \,\, P_o(Cyl^1_{\gamma_1})=P_o(Cyl^2_{\gamma_2}).
\end{equation}
But any two cylindrical functions $Cyl^1_{\gamma_1}$ and $Cyl^2_{\gamma_2}$ are $\sim_P$-equivalent iff $Cyl^1=Cyl^2$ and $\gamma_1$ is an isomorphic to $\gamma_2$ as a knot, by construction of $P_o$, which depends only on the isoknot class of the considered graph. This lets us build a map $I$, which is defined for any cylindrical function $Cyl_\gamma$ as:
\begin{equation}
  I: P_o(Cyl_\gamma) \mapsto [Cyl_\gamma]_P,
\end{equation}
which is clearly continuous with continuous inverse and turns out to be an isomorphism. We conclude that the diffeomorphism orbits of cylindrical functions are indeed isomorphic to the points in the image of $P_o$. Thus, the isomorphism $I$ allows us the interpretation of the image of $P_o$ as diffeomorphism invariant states of Loop Quantum Gravity. Since our quantum embedding $(P,i)$ is built as a pair of maps, where $P$ is the restriction of the functions in the image of $P_o$ to symmetric connections and $i$ is the assignment of a function in the image of $P_o$, we can use $I$ to interpret $i\circ P$ as diffeomorphism invariant states and thus the image of $P$ as symmetry reduced diffeomorhism invariant states.

\subsection{Inclusion of Matter}

The simplest type of matter that one can include into a Loop Quantum Gravity model is $U(1)$-Higgs matter. In this subsection, we follow the treatment of Higgs-fields from \cite{rovelli} and construct the quantum embedding for homogeneous configurations. This section is not intended to construct a realistic model, which would include a $U(1)$-gauge field and possibly fermions, but it is rather intended to explain how a model with fields other than gravity can be reduced in our framework.

The basic observation underlaying the Loop Quantization of Higgs matter is that the field operator can be exponentiated at each point. Since a zero-dimensional smearing is diffeomorphism invariant for scalar fields, we obtain point holonomies $U$, that correspond to\footnote{When taking a general scalar field, one can also consider a 0-dimensional smearing, but the point holonomies $U_x(\lambda):=\exp(i \lambda \phi(x))$ will depend on an additional parameter $\lambda$.}:
\begin{equation}
  U(x):=\exp(\phi(x) Y) \in U(1),
\end{equation}
where $Y$ denotes the generator of $U(1)$, particularly in the fundamental representation it is the $1\times 1$-matrix $(i)$. The conjugate momenta are volume forms, meaning that the three-dimensional smearing of the conjugate momenta $P$ over a region $R$ is diffeomorphism invariant:
\begin{equation}
  P(R):=\int_R d\sigma \,\,\pi(\sigma).
\end{equation}
Quantization is the based on these operators and their fundamental Poisson brackets. The cylindrical functions are functions of finitely many point-holonomies\footnote{The Ashtekar-Lewandowski measure is extended to these cylindrical functions by $\int d\mu_{AL}(A,\phi) Cyl(A,\phi)=\int d\mu_{AL}(A) d\mu_H(g_1)...d\mu_H(g_n) \psi(A,g_1,...,g_n)$, where $d\mu_H(g)$ denotes the Haar measure on $U(1)$ over the variable $g$. This defines the canonical inner product $\langle \psi_1,\psi_2 \rangle:=\int d\mu_{AL}(A,\phi) \overline{\psi_1(A,\phi)}\psi_2(A,\phi)$.}:
\begin{equation}
  Cyl(A,\phi)=\psi(h_{e_1},...,h_{e_m},U(x_1),...,U(x_n)).
\end{equation}
The Weyl-operators $W_R(-\lambda)=W^*_R(\lambda)$ that correspond to the exponentiated Poisson action of the momenta $P(R)$ on configuration variables are a unitary representation of $\mathbb R$ for each region $R$, satisfying the canonical Weyl-commutation relations, i.e. their action on the point holonomies is:
\begin{equation}
  W_R(\lambda) U(x) W_R(-\lambda)= \biggl\{
  \begin{array}{lc}
    e^{-i\lambda} U(x) & x \in R \\
    U(x) & {\textrm{otherwise}}
  \end{array}\biggr.
\end{equation}
The simplest and most natural choice is to extend the scaffold that we considered so far and to allow construct diffeomorphism fixed cylindrical functions that depend on the point holonomies at vertices of the scaffold. It is not difficult to verify that each cylindrical function can be embedded into this scaffold: (1) We already saw that any cylindrical function that depends purely on the gravitational field can be embedded into the scaffold using an extended diffeomorphism. (2) Those point holonomies that depend on a point in the graph of the gravitational field are then automatically embedded into the scaffold. However, since we did not consider a $U(1)$-gauge-field, there may be point-holonomies that do not reside on any vertex\footnote{In the absence of the $U(1)$-gauge-field, get rid of these extra point-holonomies, because we can not invoke gauge-invariance for the cylindrical functions, which forces all the point-holonomies to reside on a vertex.}. But these extra points can be mapped onto any unoccupied vertices on the scaffold using a piecewise analytic diffeomorphism, which leaves the graph invariant.

We can again specify a minimal region $R(\gamma,x_1,...,x_n)$ for any $(\gamma,x_1,...,x_n)$ into which the cylindrical functions can be surely mapped. We can thus use the same procedure as before to define the map $P_o$ by:
\begin{equation}
  P_o: Cyl \mapsto \frac{1}{|\mathcal I(\gamma,x_1,...,x_n)|} \sum_{\phi \in \mathcal I} (\phi^* Cyl),
\end{equation}
where $\mathcal I(\gamma,x_1,...,x_n)$ denotes the set of all possible topology-preserving embeddings of $(\gamma,x_1,...,x_n)$ into the restriction of the scaffold to the region $R(\gamma,x_1,...,x_n)$. Weyl-operators on 0-dimensional quasi-regions arise from the ones on 3-dimensional quasi-regions as follows: Given a point $x$ there is an open neighborhood $O(x)$ and $O(x)\setminus \{x\}$ is also open. Then $W_x:=W^*_{O(x)\setminus \{x\}}W_{O(x)}$ is a Weyl-operator supported only on $x$. We will again view these as the elementary Weyl-operators.

The procedure of defining the quantum embedding $(P,i)$ is now completely analogous to the case of pure gravity: We define $P$ for a dense set of cylindrical functions (i.e. the spin-charge-networks) by restricting these functions to their dependence on the homogeneous connection and the homogeneous Higgs field. We notice that the charge part of the spin-charge-network is given by the the assignment of an irreducible representation of $U(1)$ to all vertices and all $x_i$. Let us denote the collection of these quantum numbers by $(n_1,...,n_m)$, such that any spin-charge-network function $SCNF$ can be written in terms of a spin network function $SNF$ as:
\begin{equation}
  SCNF(A,\phi)=SNF(A)_{n_1,...,n_m} U_1^{n_1}...U_n^{n_m}.
\end{equation}
If we evaluate this function for a homogeneous Higgs-field strength $\phi(\sigma)=\phi_o$, then we obtain the $\phi_o$-dependence of $SNCF$ as:
\begin{equation}
  SNCF(A,\phi_o)=SNF(A) e^{i(n_1+...+n_m)\phi_o}.
\end{equation}
This allows us to extend the quantum embedding map $P$ to all $SNCF$:
\begin{equation}
  P: SNCF \mapsto P(SNF) e^{i\phi_o(n_1+...+n_m)}.
\end{equation}
The quantum embedding map $i$ is extended to the extension of $P$ in a very similar way:
\begin{equation}
  i: P(SNF) e^{i k \phi_o} \mapsto i(P(SNF)) U((0,0,0))^m,
\end{equation}
where $(0,0,0)$ denotes the vertex at the origin of the scaffold, which is contained in $R(\gamma,x_1,...,x_n)$ for any nonempty $(\gamma,x_1,...,x_n)$. The extension by linearity of $P$ to all cylindrical function as well as the linear extension of $i$ defines the quantum embedding map for a theory with matter. Notice that we did not have to distinguish between the application of $(P_{LRS},i_{LRS})$ and $(P_{iso},i_{iso})$, because the Higgs field transforms as a scalar under diffeomorphisms.

It is not difficult to extend the pre-Hilbert-$C^*$-module that we used for pure gravity to the matter theory: We extend the module-set, that is given by $F$ by allowing for cylindrical functions $F$ that also depend on the point holonomies. If we denote the Weyl-operator $W_R(\lambda)$, where $R$ is a the 0-dimensional quasi-region containing exactly one vertex $x$ in the scaffold by $W_x(\lambda)$, then we are able to extend the bilinear structure to the theory with matter content by first defining:
\begin{equation}
  \begin{array}{rl}
   \langle F_1, F_2 \rangle :=& F_1 \int \Pi_{e \in E(\gamma)} d\mu_H(g_e) \Pi_{x_i} d\mu_H(u_i)  \\
   & \overline{F_2(g^{-1}_{e_1}h_{e_1},...,g^{-1}_{e_n}h_{e_n},u_1^{-1}U(x_1),...,u_m^{-1}U(x_m))} \\
   &W_{S_{e_1}}(\lambda(g_{e_1}))...W_{S_{e_n}}(\lambda(g_{e_n})) W_{x_1}(\lambda(u_1))...W_{x_m}(\lambda(u_m))
  \end{array}
\end{equation}
and then again averaging over all embeddings $\phi\in \mathcal I(\gamma,x_1,...,x_m)$ of $\gamma,x_1,...,x_m$ into the region $R(\gamma,x_1,...,x_m)$, which yields:
\begin{equation}
  \langle F_1 , F_2 \rangle_A:=\frac{1}{|\mathcal I(\gamma,x_1,...,x_m)|} \sum_{\phi \in \mathcal I} \langle F_1, F_2 \rangle,
\end{equation}
which spans a dense set in the Weyl-algebra of the matter theory.

The module set for the embedded matter theory is spanned by $(n,m,k):=e^{i(ma+nc+k\phi_o)}$. The bilinear structure for the embedded module can be calculated rather easily. For the LRS-matter model it becomes:
\begin{equation}
  \langle (n_1,m_1,k_1) , (n_2,m_2,k_2) \rangle_{LRS} (n,m,k) = \delta_{n_2,n}\delta_{m_2,m}\delta_{k_2,k} (n_1,m_1,k_1).
\end{equation}
The isotropic structure arises as a special case e.g. by setting all $n=0$. This structure allows us to induce the ground state for the $LRS$-matter model:
\begin{equation}
  \begin{array}{rl}
    \omega_{LRS}(\langle (n_1,m_1,k_1) , (n_2,m_2,k_2) \rangle_{LRS}):=&\omega(\langle i((n_1,m_1,k_1)) , i((n_2,m_2,k_2)) \rangle_A)\\
     & = \delta_{n_1,o}\delta{n_2,o}\delta_{m_1,o}\delta_{m_2,o}\delta_{k_1,o}\delta_{k_2,o},
  \end{array}
\end{equation}
which is calculated using exactly the same arguments as in the matter free case and by using that $U((0,0,0))^{n}$ is normalized and orthogonal to $U((0,0,0))^m$ for $n \ne m$. We immediately see that this model is quantum mechanics of a particle on the 3-torus $\mathbb T^3$. The isotropic model arises as the spacial case for $m_1=m_2=0$ and is thus quantum mechanics on $\mathbb T^2$.

Using the same arguments as in the previous subsection, we can identify the states in the image of $i$ as diffeomorphism invariant states and we are thus able to call our reduced model diffeomorphism invariant. The construction of gauge-invariant matter states is however more involved and we will postpone it to future work, because we did not introduce the $U(1)$-gauge-field that the Higgs-field couples to.

\subsection{Dynamics}\label{dynamics}

This section is concerned with the treatment of the Hamilton constraint in our construction. We will find out that the quantum embedding $(P,i)$ needs significant improvement before it can be applied to this framework. We will sketch these improvements, but to carry them out needs significantly more work and is beyond the scope of this paper. Thus, we are not really defining a dynamics for the cosmological models in this section.

The idea that we used to construct the gauge invariant quantum embedding $(P,i)$ system was to restrict the domain of $P$ to gauge invariant states in the full theory and to construct $i$ such that it takes values in the gauge invariant states of the full theory. There is no reason to treat the Hamilton constraint any differently. But this leads to an astonishing result: Consider the LRS-model and let us restrict $P_{LRS}$ to the solution space of the Hamilton constrints\footnote{The Hamilton constraints are given as a set of constraints, that are labeled by different lapse functions, but we will simply refer to it as the Hamilton constraint.}.

The important observation is that powers of traces of holonomies over closed loops are sufficient to construct the general dependence of a gauge-invariant state on the homogeneous connection. But due to homogeneity we can choose these loops to be completely separated. This means for each gauge-invariant continuous function (defined through a function with graph in the scaffold) on the reduced connection there is a cylindrical function that depends on a graph with only bi-valent vertices. These cylindrical functions are annihilated by all known versions of the Hamilton constraint operator that are discussed in literature. Thus, the restriction of the gauge-invariant $P_{LRS}$ to solutions of the Hamilton constraint does not reduce its image. Similarly, we can choose $i_{LRS}$ to take values in cylindrical functions with only bi-valent vertices. This means that the Hamilton constraint is satisfied by all gauge invariant $LRS$-observables.

This is in sharp contrast to the classical theory, where the Hamilton constraint acts nontrivially on locally rotationally symmetric homogeneous observables. This huge solution space poses obvious problems on the classical limit of the embedded theory and thus on the consistency and physical interpretation of this approach all together. So, how odes this enlargement of the solution space come about?

The first guess for an answer would probably rest on the observation that the states that we use to construct in the image of $i$ correspond to degenerate geometries with no volume and disconnected area segments. Then one can assert that a construction that relies solely on these states should be rejected on physical grounds. So the question is: Is there a mathematical reason that has a reasonable physical interpretation to reject a construction using these ''pathological'' states?

In section \ref{intro} we discussed the superficial homogeneity of the construction of $P$. And this is where the problem lies: It is not at all clear how the gauge fixing of the diffeomorphisms has to be constructed such that a superficial homogeneity there implies geometrical homogeneity in a suitable coarse graining of the geometric observables.

Thus, the solution to the problem posed here is as follows: Instead of using ''superficial homogeneity'', we have to construct geometrical homogeneity. This means instead of building equivalence classes of functions of the connection and try to achieve geometrical homogeneity through the construction of a suitable gauge fixing of the diffeomorphisms, we should build a net of geometrical observables, that are homogeneous in some direct sense, and build equivalence classes through the expectation values of these operators. We could proceed as follows:
\begin{enumerate}
  \item We reuse the scaffold and the embedding of $Cyl_\gamma$ into $R_\gamma$ on the scaffold.
  \item For each cube $R$, we define a set of geometrical operators that are sufficient to distinguish between all homogeneous geometries and that are compatible as we move from $R_1$ to a larger cube $R_2$. These could be areas of the dual lattice to the restriction of the scaffold to $R_\gamma$. Homogeneity (or any other symmetry reduction) then implies relations between these areas and we call a state {\bf geometrically homogeneous}, if these implied relations are satisfied by the expectation values of the respective area operators.
  \item We construct $P$ by building equivalence classes w.r.t. this set of operators and complete the construction of the quantum embedding by assigning a geometrically homogeneous state $i(.)$ to each equivalence class.
\end{enumerate}
The reason for first trying to construct this map from area operators rather than volume operators is the significantly larger complexity to the calculation of a volume compared to an area\footnote{As shown in \cite{dqg}, one can construct a volume operator form area operators alone, so the restriction to a set of areas seems not to loose any geometric information about the state.}. The explicit construction is even for pure area operators rather involved and far beyond the scope of this paper. We will only discuss what we mean by this net of geometrical operators in the context of Bianchi I cosmology:

Suppose we are given a cube centered around the origin of the scaffold. Let us first consider the regular lattice in the scaffold. To each edge $e^i_{abc}$ in the regular lattice, we associate a surface $S^i_{abc}$, that is a unit square and that intersects $e^i_{abc}$ perpendicularly:
\begin{equation}
  S^i_{abc}:=\{\phi(t_1 e^j+t_2 e^k+e^i_{abc}(1/2)):-1/2 \le t_i \le 1/2\},
\end{equation}
where $e^i_{abc}(1/2)$ denotes the center of $e^i_{abc}$ i.e. the chart $U,\phi$ and $\hat e^j$ and $\hat e^k$ are two mutually orthogonal unit vectors that are perpendicular to $\dot{e}^i_{abc}$ in this chart. We want to use the area operators on these surfaces to resolve different geometries. Let us assume that we want to resolve homogeneous isotropic geometries. homogeneity means for these small surfaces that the area is invariant under translation by a lattice vector, i.e. inside the cube, we want that for any pair of surfaces $S^i_{abc},S^i_{a+n_1,b+n_2,c+n_3}$:
\begin{equation}
  A(S^i_{abc}) = A(S^i_{a+n_1,b+n_2,c+n_3})
\end{equation}
and isotropy implies that
\begin{equation}
  A(S^i_{abc}) = A(S^j_{abc})
\end{equation}
for any two directions $i$ and $j$. This is a geometric constraint that we need to keep in mind, when we construct the quantum embedding $(P,i)$, i.e. the states in image of the map $i$ should satisfy such a set of constraints through their expectation values, e.g.:
\begin{equation}
  \langle Cyl, A(S^i_{abc}) Cyl \rangle = \langle Cyl,A(S^i_{a+n_1,b+n_2,c+n_3}) Cyl \rangle.
\end{equation}
This set of constraints leads us naturally to the construction of cylindrical functions that depend on the lattice in an even way, i.e. states of the form:
\begin{equation}
   Cyl(A)=\Pi_{i,abc\in R} (Tr(h_{L^i_{abc}}(A)))^n,
\end{equation}
where $L^i_{abc}$ denotes a right handed loop based on the point $(a,b,c) \in R$ that encloses a surface which is normal to the direction $i$ in the scaffold chart. These functions satisfy the geometric constraints naturally, but it is not clear how to construct solutions to the Hamilton constraint from these, which is due to the fact that all vertices in the respective graphs are six-valent. These difficulties are the reason, why we have to postpone the discussion of dynamics to future work.

\section{Comparison with Standard Loop Quantum Cosmology}

We have already mentioned differences between our approach outlined in this paper and Loop Quantum Cosmology as it is presented by \cite{abl}. The apparent difference is the different treatment of the diffeomorphisms, which we treat as pure gauge in the full theory and we preform the symmetry reduction only after this gauge freedom is fixed. In this sense, we view the diffeomorphism invariance of LQG as more fundamental than the classical symmetry reduction, which forces us to preform the symmetry reduction in this way.

The configuration variables and their Hilbert-space representation of Loop Quantum Cosmology can be viewed as being constructed using the following quantum embedding $(P_{LQC},i_{LQC})$:

Fix a chart $(U,\phi)$ and fix one base-point $x_o$ and one direction $v^i \partial_i$ in this chart. Map any cylindrical functions $Cyl_{\gamma}$ into a function on $SU(2)$, that depends on $|E(\gamma)|$ copies of $SU(2)$, but only on straight edges $\{s_i(\gamma)\}$, whose  Euklidean length is the Euklidean length of each edge $e_i$ in the chart and which start at $x_o$ and have the tangent vector $v^i \partial_i$. Then evaluate these functions on homogeneous, isotropic connections. This defines the map $P_{LQC}$ as
\begin{equation}
  \begin{array}{rl}
    P_{LQC}: & \left( A \mapsto (Cyl_{\{e_1,...,e_n\}}(h_{e_1}(A),...,h_{e_n})\right) \\
    \mapsto & \left( c \mapsto Cyl_{s_1,...,s_n}(h_{s_1}(A(c)),...,h_{s_n}(A(c)))\right),
  \end{array}
\end{equation}
where $A(c)$ denoted the dependence of the isotropic connection on the parameter $c$. Choosing a diagonal gauge and the direction $\partial_3$ as well as the standard $\tau$ matrices yields a dependence on the holonomies:
\begin{equation}
  h_s = \biggl(
    \begin{array}{cc}
      e^{i l_s/2 c} & 0 \\
      0 & e^{-i l_s/2 c}
    \end{array}
  \biggr),
\end{equation}
where $l_s$ denotes the Euklidean length of $s$ in the chosen chart. The image of a spin network function, which is a sum of products of matrix elements of these elementary holonomies is:
\begin{equation}
  P(SNF_\gamma)(c) = \sum_{j,n} \xi_{j,n} e^{i n l_{s_j}/2 c},
\end{equation}
which spans the continuous functions on $\mathbb R_{Bohr}$. We define the map $i_{LQC}$ directly on these functions:
\begin{equation}
  i(c \mapsto \sum_{j,n} \xi_{j,n} e^{i n l_{s_j}/2 c}):= \sum_{j,n} \xi_{j,n} (h_{n l_{s_j}})_{11},
\end{equation}
where $n l_{s_j}$ denotes the edge $\{(0,0,t): 0 \le t \le n l_{s_j}\}$. Using the Schr\"odinger ground state on Loop Quantum Gravity (equation \ref{schr-fun}), we can induce the ground ground state of $\mathbb R_{Bohr}$.

An interesting feature of Loop Quantum Cosmology is the over-countable number of eigenvalues of the flux operators. Let us now show, that we do not obtain this phenomenon, if we follow the quantum embedding procedure using $(P_{LRS},i_{LRS})$. For this purpose, we will consider a flux $E(S_3)$ through a unit square $S_3$ centered at $(0,0,\frac 1 2)$ that is normal to the 3-axis. This Flux operator acts on the scaffold as the flux through the a  0-dimensional quasi-surface $x_3=(0,0,\frac 1 2)$ with the respective orientation. This vertex has to be introduced by hand into the scaffold by splitting the respective edge. This flux arises as the generator of the Weyl-operator $W_\lambda(x_3)$, where $\lambda$ is constant and takes values in $su(2)$. Using an approximate identity for the embedded reduced system $id=\lim_{i\in \mathcal I}\sum_{j} \langle i(f^i_j), i(f^i_j) \rangle_A$, we obtain the action of the induced Weyl-operator:
\begin{equation}
  W_{red}(\lambda,x_3):=\lim_{i\in \mathcal I}\sum_{j} \langle i(f^i_j), W_{-\lambda}(x_3) i(f^i_j) \rangle_A.
\end{equation}
However, we see that the action of $W_{-\lambda}(x_3)$ on $i(f^i_j)$ is generated by the action of $E(x_3)$. The action of $E(x_3)$ on the holonomies $h_{s_i}$ is however:
\begin{equation}
  E(x_3) h_{l_i}(A) = \tau^3 h_{s_i}(A),
\end{equation}
and we obtain that the spectrum of this flux operator consists of only two points and is independent of the Euklidean length $l_i$ and of course $0$ for $Cyl(A)=1$. The same argument holds obviously for any other flux operator, too.\footnote{It might be bothersome to the reader to have such an action of the flux operators. One can construct $i$ to be $i:e^{i l / 2 c}\mapsto (h_{l-[l]})_{11} \rho^n(h_{[l]})_{11}$, where $[l]$ denotes the integral part of $l$. The induced flux operator $E(S_3)$ then takes an integral eigenvalue spectrum, which is again independent of the non integral part $l-[l]$ for non integral real numbers $l$.}

An important difference to our approach is that the map $P_{LQG}$ does not respect the knot class of a graph. This means that on the one hand $P_{LQG}$ can map $Cyl_{\gamma_1}$ and $Cyl_{\gamma_2}$ into the same $P(Cyl)(c)$ even if $\gamma_1$ and $\gamma_2$ are not in the same knot class, but on the other hand, it does map $Cyl_{\gamma}$ and $Cyl_{\phi(\gamma)}$ into two different reduced functions $P(Cyl)(c)$, as can be seen by applying a simple diffeomorphism, that takes the form
\begin{equation}
  \phi:(x_1,x_2,x_3)\mapsto(a x_1,a x_2,a x_3) \label{diff}
\end{equation}
in the cahrt $U,\phi$, whenever $a \ne 0,1$. This means that $P_{LQC}$ does not respect knot classes, which is a consequence of the different treatment of the diffeomorphisms in full LQG compared to the construction presented in this paper.

\subsubsection*{Nonembedability of Loop Quantum Cosmology}

Brunnemann and Fleischhack showed in \cite{fleischhack-brunnemann} that the kinematical configuration space of standard LQC is not the restriction of the kinematic configuration space of LQG, by calculating the dependence of holonomies along spiral curves on isotropic connections, which means that kinematic LQC is not embedded into LQG. It is not obvious which enlargement of the configuration space of LQG is the restriction of kinematic LQG to isotropic connections. By fixing the diffeomorphisms however, we are able to explicitly calculate the dependence of diffeomorphism fixed cylindrical functions on the isotropic connection and thus avoid the problem of constructing the kinematic restriction of cylindrical functions.

Using the graph groupoid to represent the action of the diffeomorphisms as presented in this paper, we removed the super-selection sectors that occur in the standard representation of LQC at the level of diffeomorphism invariant states. These ''extra states''that appear in Loop Quantum Cosmology however can be removed by considering orbits of the one-parameter family $\mathcal D_o$ of diffeomorphisms of the form \ref{diff}, i.e.:
\begin{equation}
  \mathcal D_o=\{\phi:(x_1,x_2,x_3)\mapsto(a x_1,a x_2,a x_3); a \in \mathbb R^+\}\subset \mathcal D
\end{equation}
This has the immediate consequence that this particular quantum embedding $(P_{LQC},i_{LQC})$ can not be embedded into the diffeomorphism invariant theory of Loop Quantum Gravity. Due to the diffeomorphism invariance of the construction in this paper, we have a quantum embedding of our reduced theory into diffeomorphism invariant LQG.

Let us now focus on the differences in the philosophy between our approach and Loop Quantum Cosmology.

\subsubsection*{Treatment of Observables and States}

We reduced the full theory of Loop Quantum Gravity using a procedure that is reminiscent of constructing equivalence classes of states and the fixing a representative in each equivalence class. This has the immediate consequence that the states in our reduced model do have an interpretation as states of the full theory. The states of LQC need a physical interpretation and is not induced by the relation of Loop Quantum Cosmology and Loop Quantum Gravity itself; see \cite{engle} for a discussion of the physical interpretation of the states of Loop Quantum Cosmology.

The correspondence that we constructed through $(P_{LQC},i_{LQC})$ is only used to construct the Hilbert-space representation of the configuration observables, but not used to construct the reduced algebra in particular not the Weyl-operators for the reduced theory. Loop Quantum Cosmology induces the quantization of the symmetric fluxes from full Loop Quantum Gravity and thus needs additional input from LQG for the quantization of composite operators. Using the Weyl-operators in the approach presented in this paper provides a prescription for the quantization of composite operators.

We saw by constructing a quantum embedding $(P_{LQC},i_{LQC})$, that the sates of LQC can be constructed in a very similar fashion to the symmetry reduction used in this paper. It seems therefore likely that the selection of a particular super-selection sector in LQC may be constructed using a geometrically homogeneous quantum embedding and the approach presented in this paper. One could the identify these super-selection sectors of LQC with symmetric sectors of diffeomorphism invariant full Loop Quantum Gravity.

\subsubsection*{Treatment of Constraints and Operators}

Since our construction takes states of the full theory of LQG, we are able to impose the full set od constraints of the full theory on our construction. Using this procedure ensures that the gauge-invariant states of our reduced theory correspond to gauge invariant states in the full theory.

In the construction of standard Loop Quantum Cosmology however, one imposes a set of constraints that is classically induced on the reduced classical phase space. Although this procedure is classically equivalent, we see that the application to the quantum theory yields different reduced theories. For example: The diffeomorphism constraint for isotropic cosmology is classically empty and this is how it is implemented in standard Loop Quantum Cosmology. However, as we saw by explicitly constructing the set of diffeormorphisms $\mathcal D_o$, one can not distinguish between graphs consisting of one edge and a graph that consists again of a single edge but of different Euklidean length.

There is a clear correspondence between the classical cosmological observables and the operators in Loop Quantum Cosmology. The interpretation of these operators in terms of the full theory is done by comparing the classical limits of cosmological operators and full operators. As a consequence, it is not possible to ensure that a cosmological operator has a spectrum that is a subset of the spectrum of the corresponding full operator, as we saw explicitly with the flux operators.

The situation in our approach is the other way around: We do have a direct correspondence between the operators in the cosmological sector and the one in the full theory, but we do not have a clear correspondence between the classical cosmological observables and the operators of our reduced model. The interpretation of the operators of our model is constructed by taking the classical limit of the corresponding operators in the full theory and considering the cosmological sector of this classical limit. The construction of this correspondence is one of the main open problems of our approach and is not completely resolved until the classical limit of Loop Quantum Gravity can be constructed without too many assumptions.

\section{Conclusion}

The intention of this work is to contribute to the understanding of the relation of Loop Quantum Gravity and Loop Quantum Cosmology. We have expressed the view that the construction presented in this paper could be used to interpret super-selection sectors of standard LQC as cosmological sectors of diffeomorphism invariant full LQG. The long term hope is to establish a firm basis for this speculation and thus provide an embedding of the super-selection sectors of LQC as cosmological sectors of the diffeomorphism invariant sector of full theory despite the non-embedability of the kinematic configuration spaces.

We started our investigation by considering classical cosmology as an embedded subsystem of General Relativity and we used this relation to construct a quantum embedding of cosmological models into Loop Quantum Gravity. The construction of the quantum embedding needed a gauge fixing of the diffeomorphisms, which we achieved by assigning exactly one representative function to each knot class of cylindrical functions. This amounted to the construction of a combinatorial theory, whose cylindrical functions are labeled by equivalence classes of graphs on a combinatorial groupoid. This allowed us to construct the quantum embedding on a theory defined on the combinatorial groupoid:

We embedded the combinatorial groupoid into the path groupoid as a subgroupoid, which defiens a ''scaffold''. The quantum embedding on cylindrical functions that depend only on graphs in this ''scaffold'' are defined through the pair of linear maps that ($P$) returns the dependence on the degrees of freedom of the homogeneous connection and ($i$) assigns functions depending on degrees of freedom of the homogeneous connection a representative cylindrical function that depends on a graph whose edges are all elements of the embedded combinatorial groupoid. The diffeomorphism invariance was used to extend the domain of $P$ to all cylindrical functions.

Next we used a Schr\"odinger type Hilbert-$C^*$-module for the combinatorial theory to construct the reduced theory. It is possible to use a Hilbert-$C^*$-module for the combinatorial theory due to diffeomorphism invariance of observables in Loop Quantum Gravity, such that a gauge-fixing of the diffeomorphisms maps cylindrical functions on any graph into cylindrical functions on the scaffold. This reduced theory turned out to be quantum mechanics on a 2-torus for a locally rotationally symmetric cosmological model and quantum mechanics on a circle for an isotropic cosmological model.

The reduced theory is, by our construction, already diffeomrophism invariant, but we still had to impose the Gauss-constraint. This means that both the domain of the quantum embedding map $P$ and the range of $i$ had to be restricted to gauge-invariant cylindrical functions. The reduced quantum theory constructed with this embedding is a subsystem of the gauge variant theory, and is gauge invariant in the sense that it relates the reduced theory only to gauge invariant operators in the full theory.

We showed how matter can be treated in this framework by demonstrating it on the example of a $U(1)$-Higgs field. Since the Higgs field is only supported on vertices, we where again able to gauge fix the diffeomorphisms and to define the map $P$ as the extension of the previous map $P$ that also assigns the dependence on the homogeneous degree of freedom of the Higgs field to each cylindrical function. The map $i$ is similarly extended by assigning a representative cylindrical function, that depends on the Higgs degrees of freedom in a given way. The induced quantum theory for the locally rotationally symmetric model turns out to be quantum mechanics on a 3-torus, whereas the isotropic model turns out to be equivalent to quantum mechanics on the $2$-torus.

The discussion of the dynamics of the reduced system revealed that the reduced system relies heavily on states that correspond to degenerate geometries, i.e. geometries whose large scale structure is not three-dimensional. These states that are negligible in the full theory, provide many extra solutions to the Hamilton constraints, which spoil the discussion of dynamics of the reduced models. We thus proposed a geometric version of the quantum embedding $(P,i)$, and explained why the dynamics induced through a geometric model would avoid these problems.

Besides a deeper comparison of our approach of ''constructing cosmological sectors in full Loop Quantum Gravity'' with the standard Loop Quantum Cosmology, there are many loose ends to this work and we want to just outline some of the many future works that seem interesting from the present work and we want to mention only three that are already under investigation:

(1) There is of course the problem of the construction of a geometric quantum embedding as proposed in section (4.5), for which we need to calculate the reduced quantum algebra and the induced representation, before we are able to discuss the induced dynamics of cosmological models. A set of states \cite{dqg} that are exact geometrically homogeneous states where found in the time between working on this paper and posting it.

(2) One needs to become able to define ''small perturbations'' of the homogeneous models, such that one is enabled to define a ''hierarchy of more and more negligible perturbations''. This needs an extension of our formalism to an approximate quantum embedding $P_\lambda,i_\lambda$, where the powers of a supposedly small parameter $\lambda$ denote the order of the perturbations that are considered. This will enable us to induced a family of quantum algebras and induced representations, which will allow for perturbative ans\"atzae. This may be achieved using the states of \cite{dqg}, because these states are states on the full observable algebra, which carry a natural measure for the size of perturbations, due to a measurable classical background geometry in these states.

\subsection*{Aknowledgements}

This work was in part supported by the Deutsche Forschungsgemeinschaft. I wish to thank Martin Bojowald for a careful reading of the script, useful comments and a clarifying discussion about the current research on the relation between Loop Quantum Gravity and Loop Quantum Cosmology, particularly the construction of momentum operators as averaged momentum operators of the full theory.

\end{document}